\documentclass[preprint,aps,amsmath,nofootinbib,floatfix]{revtex4}
\usepackage{graphicx}
\usepackage{multirow}
\usepackage{array}
\usepackage{color}
\usepackage{comment}
\usepackage{mathrsfs}
\usepackage{epstopdf}
 \usepackage{booktabs}  
 \usepackage{natbib}

\makeatletter

\newcommand{\Rmnum}[1]{\expandafter\@slowromancap\romannumeral #1@}
\makeatother
\newcommand{\met}{\ensuremath{{\not\mathrel{E}}_T}}

\def\sL{\ensuremath{\tilde t_L}}
\def\sR{\ensuremath{\tilde t_R}}

\begin{document}

\title{ {Higgs and $Z$ Assisted Stop Searches at Hadron Colliders}}
\author{
         Shufang Su$^{1}$\footnote{shufang@email.arizona.edu},
         Huanian Zhang$^{2}$\footnote{fantasyzhn@email.arizona.edu}}

\affiliation{
	$^{1}$ Department of Physics, University of Arizona, Tucson, Arizona  85721}
\affiliation{
$^{2}$ Steward Observatory, University of Arizona, Tucson, Arizona 85721}

\begin{abstract}

	Current searches for the light top squark (stop) mostly focus on the decay channels of $\tilde{t} \rightarrow t \chi_1^0$ or $\tilde{t} \rightarrow b \chi_1^\pm \rightarrow bW \chi_1^0$, leading to $t\bar{t}/bbWW+\met$ final states for   stop pair productions at the LHC.  However, in   supersymmetric scenarios with light neutralinos and charginos other than the neutralino lightest supersymmetric particle (LSP), more than one   decay mode of the stop could be dominant.  While those new  decay modes could significantly weaken the current stop search limits at the LHC, they also offer alternative discovery channels for stop searches.  In this paper, we studied the scenario with light Higgsino next-to-LSPs (NLSPs) and Bino LSP.    The light stop  decays primarily  via $\tilde t_1 \to t \chi_2^0/\chi_3^0$,  with the neutralinos subsequent decaying to a $Z$ boson or a Higgs boson: $\chi_2^0/\chi^0_3 \to \chi_1^0 h/Z$.  Pair production of light stops at the LHC leads to final states of $t \bar t hh\met$, $t \bar t hZ\met$ or $t \bar t ZZ\met$.     We consider three signal regions: one charged lepton (1$\ell$), two opposite sign charged leptons (2 OS $\ell$) and at least three charged leptons ($ \ge  3 \ell$).  We found that the 1$\ell$ signal region of channel $t \bar t hZ\met$ has the best reach sensitivity for light stop searches.   For 14 TeV LHC with 300 ${\rm fb}^{-1}$ integrated luminosity, a stop mass up to 900 GeV can be discovered at 5$\sigma$ significance, or up to  1050 GeV can be excluded at 95\% C.L. Combining all three decay channels for $1 \ell$  signal region extends the reach for about 100$-$150 GeV.  We also studied the stop reach at the 100 TeV $pp$ collider with 3 ${\rm ab}^{-1}$ luminosity, with discovery and exclusion reach being 6 TeV and 7 TeV, respectively.  	
\end{abstract}

\maketitle

\section{Introduction}

The milestone discovery of a light Standard Model (SM)-like Higgs boson at the Large Hadron Collider (LHC) \cite{Aad:2012tfa,Chatrchyan:2012xdj} calls for  the new physics beyond the SM to solve the ``Hierarchy problem''~\cite{Weinberg:1975gm}. Among various   new physics beyond the Standard Model, Supersymmetry (SUSY) remains to be one of  the most attractive candidates because of the elegant solution to the ``Hierarchy problem'' and the accommodation of the light SM-like Higgs. In the supersymmetric models, the third generation scalar tops (stop) might be the most relevant ones given the large top Yukawa coupling to the Higgs sector.   Searching for the heavy top partners is one of the primary goals of the LHC to solve the puzzle of electroweak symmetry breaking and the stabilization of the weak scale.   The masses for stops are constrained to be less than about a few TeV to avoid extra fine-tuning to the light Higgs mass.  There are two scalar tops in the Minimal Supersymmetric Standard Model (MSSM): $\tilde{t}_L$ and $\tilde{t}_R$, which are the superpartners of the left- and right-handed top quarks, respectively.   To provide large enough loop corrections to the tree-level Higgs mass ($m_{h}^{\rm tree}\leq m_Z$), a large left-right mixing between $\tilde{t}_L$ and $\tilde{t}_R$ is typically needed,   leading to two mass eigenstates, $\tilde t_1$ and $\tilde t_2$,  with relatively large mass splittings.  One of the stops can be as light as a few hundred GeV, leaving the LHC an ideal place to search for those relatively light stops.
 
There are many ongoing searches for the stops by the ATLAS and CMS groups~\cite{Aaboud:2017aeu,Aaboud:2017ayj,Aaboud:2017wqg,Aaboud:2017nfd,Aaboud:2017dmy,Sirunyan:2017xse,Sirunyan:2017leh,Sirunyan:2017cwe,Sirunyan:2017kqq,Sirunyan:2017pjw,CMS-PAS-SUS-16-052, Sirunyan:2018iwl,Sirunyan:2017kiw,Sirunyan:2017wif,Aaboud:2017ejf,Khachatryan:2014doa}   and most of the searches focus on the following decay modes: $\tilde t_1 \to t \chi_1^0$ and $\tilde t_1 \to b \chi_1^\pm \to b W \chi_1^0$,  assuming a 100\% decay branching fraction into those two channels.    The current experimental search limits from those two channels exclude the stop mass up to 1120 GeV for a very light Lightest Supersymmetric Particle (LSP) $\chi_1^0$~\cite{Aaboud:2017aeu,Aaboud:2017ayj,Aaboud:2017wqg,Aaboud:2017nfd,Aaboud:2017dmy,Sirunyan:2017xse,Sirunyan:2017leh,Sirunyan:2017cwe,Sirunyan:2017kqq,Sirunyan:2017pjw}.  In cases where the mass spliting between  the stops and  the LSP is very small, the stop mass up to about 580 GeV is excluded~\cite{CMS-PAS-SUS-16-052, Sirunyan:2018iwl,Sirunyan:2017kiw,Sirunyan:2017wif,Aaboud:2017aeu}  for decay channels of $\tilde t_1 \to c \chi_1^0$ and $\tilde t_1 \to b f f^\prime \chi_1^0$. There are also direct searches for the heavier scalar top by the ATLAS and CMS groups using   the decay channel of $\tilde t_2 \to  \tilde t_1 h/Z$ ~\cite{Khachatryan:2014doa, CMS-PAS-SUS-13-021}, with $\tilde t_1$ further decaying to a top quark and an LSP near the top quark threshold. The heavier stop mass is excluded up to about 800 GeV at 95\% C.L. for final states with a $Z$ and/or $h$, assuming a 100\% decay branching fraction.

The current light stop searches  considered  a 100\% decay branching fraction of stops decaying into particular search channels for simplicity.   However, in realistic MSSM, there are typically more than one    decay modes open, depending on the mass spectrum of neutralinos and charginos, which significantly weakens the current search limits \cite{Eckel:2014wta,Han:2015tua}. The   scenario we consider in this work is  Higgsino-like Next-to-LSPs (NLSPs) and a Bino-like LSP with mass hierarchy $M_1 < \mu < M_{3SQ} \ll M_2$. The lighter stop dominantly decays via $\tilde t_1 \to t \chi_2^0 / \chi_3^0$ given the large ${\rm SU}(2)_L$ gauge coupling and large top Yukawa coupling of a mostly left-handed $\tilde t_1$,  with neutralinos subsequent decaying to a gauge boson or a Higgs boson $\chi_2^0 / \chi_3^0 \to \chi_1^0 h/Z$, leading to $t\bar{t}hh\met$, $t\bar{t}ZZ\met$ or $t\bar{t}hZ\met$ final states for the stop pair production at the LHC. Given the relatively clean  final states containing at least one lepton at the LHC, our search   regions are characterized by the charged leptons: 1 $\ell$ signal region   with exact one lepton ($e$ or $\mu$), 2 OS $\ell$  signal region with  exact two opposite-sign (OS)  leptons, and $\ge 3 \ell$ signal region with at least three leptons.  

The rest of the paper is organized as follows. In Section~\ref{sec:MSSM_stop}, we briefly review the stop sector in the MSSM, introduce the mass and mixing parameters, and explore the stop decay in different scenarios. In Section~\ref{sec:limit}, we summarize the current LHC search limits on stop search by both ATLAS and CMS collaborations, and validate our simulation with the CMS study of $\tilde{t}_2$~\cite{Khachatryan:2014doa}, which has the same final states as our process.  We also recast the CMS results in $m_{\tilde{t}_1}$ vs. $m_{\chi_1^0}$ plane.     In Section~\ref{sec:analysesof14}, we perform a detailed collider analysis of stop search sensitivity in the three signal regions at the $\sqrt{s} =$ 14 TeV LHC. In Section~\ref{sec:analysesof100}, we extend our analyses to   the future $\sqrt{s} =$ 100 TeV $pp$ machine. In Section~\ref{sec:conclusion}, we  conclude.

\section{MSSM stop sector}
\label{sec:MSSM_stop}

We work in the framework of the MSSM and focus primarily on the third generation squark sector, with relatively light Higgsino-like NLSPs (a small $|\mu|$)  and a Bino-like LSP (a small $M_1$).   Other SUSY particles including the Winos, gluinos, sleptons, and the first and second generation squarks are assumed to be heavy and decoupled to be 2 TeV. We also decouple the non-SM heavy Higgses by setting $m_A$ to be 2 TeV.

The gauge eigenstates of the third generation squarks are $(\tilde t_L,\tilde b_L), \tilde t_R$ and $\tilde b_R$,  with    $(\tilde t_L,\tilde b_L)$ forming a  $ {\rm SU(2)}_L$ doublet with a soft SUSY breaking mass $M_{3SQ}$,    $\tilde t_R$ and $\tilde b_R$ being  $ {\rm SU(2)}_L$ singlets with soft breaking masses $M_{3SU}$, and $M_{3SD}$, respectively.  The mass matrix of the stop sector is \cite{Martin:1997ns,Chung:2003fi}
 \begin{equation}
  \bf{m_{\tilde t}^2} =
  \begin{pmatrix}
    M_{3SQ}^2 + m_t^2 + \Delta_{\tilde u_L} & m_t \tilde A_t \\
    m_t \tilde A_t & M_{3SU}^2 + m_t^2 + \Delta_{\tilde u_R}
  \end{pmatrix},
\end{equation}
where the   $\Delta_{\tilde u_L}$ and $\Delta_{\tilde u_R}$ terms come from the D-term contribution in the MSSM, which are to the order of $m_Z^2$.
The off-diagonal left-right mixing term $\tilde A_t$ is given by:
\begin{equation}
\tilde A_t = A_t - \mu/\tan{\beta},
\end{equation}
with $A_t$ representing the trilinear coupling,  $\tan\beta=\langle H_u^0 \rangle/\langle H_d^0 \rangle$ being the ratio of the  vacuum expectation values of two Higgs fields $H_u^0$ and $H_d^0$ in the MSSM.

The stop mass matrix can be diagonalized with mixing angle $\theta_t$: 
\begin{equation}
\begin{pmatrix}
\tilde{t}_1 \\ \tilde{t}_2
\end{pmatrix} =
	\begin{pmatrix}
	\cos{\theta_t} & -\sin{\theta_t}\\
	\sin{\theta_t} &  \cos{\theta_t}
	\end{pmatrix}
\begin{pmatrix}
\sL \\ \sR
\end{pmatrix},
\end{equation}
resulting in two mass eigenstates $\tilde{t}_1$ and $\tilde{t}_2$, with convention $m_{\tilde{t}_1} < m_{\tilde{t}_2}$.  For $M_{3SQ}< (>) M_{3SU}$, $\tilde{t}_1$ is mostly left-handed  (right-handed), while for $M_{3SQ} \sim M_{3SU}$, $\tilde{t}_{1,2}$ could be mixtures of $\tilde t_L$ and $\tilde t_R$.

Given the large top Yukawa coupling, the stop sector provides the dominant contribution to the radiative corrections of the SM-like Higgs mass in the MSSM.  For $M_{3SQ}= M_{3SU}= M_{SUSY}$, the correction to the SM-like Higgs mass squared is~\cite{Carena:1995bx, Carena:1995wu}:
\begin{equation}
\delta m_h^2 = \frac{3}{4\pi^2} y_t^2 m_t^2 \sin^2{\beta}
	\left(	\log{\frac{M_{SUSY}^2}{m_t^2}} + \frac{\tilde A_t^2}{M_{SUSY}^2} \left( 1 - \frac{\tilde A_t^2}{12 M_{SUSY}^2}   \right)   \right).
\end{equation}
In the minimal mixing case with $\tilde{A}_t=0$, a large $M_{SUSY}$ around 5$\sim$10 TeV is needed to guarantee a SM-like Higgs mass $\sim$ 125 GeV.  In the maximal mixing case with $\tilde{A}_t=\sqrt{6}M_{SUSY}$, a relatively small $M_{SUSY} \sim$ TeV  can be accommodated given the additional contribution from the $\tilde{A}_t$ term.  In the general MSSM where  $M_{3SQ}^2\neq M_{3SU}^2$, the light stop $\tilde{t}_1$  as light as 200 GeV is still consistent with a SM-like Higgs mass around 125 GeV.  A large mass splitting between the stop mass eigenstates, however,  is typically needed, resulting in $m_{\tilde{t}_2} \gtrsim 500$ GeV in general \cite{Christensen:2012ei, Carena:2011aa}.

In the scenario of Higgsino-like NLSPs and a Bino-like LSP, the two neutralinos $\chi_2^0$, $\chi_3^0$ and charginos $\chi_1^\pm$ are nearly degenerate, leading to  almost undistinguishable collider signals.   To illustrate the MSSM mass parameters and the corresponding mass spectrum, we showed one benchmark point in Table \ref{table:MassParameters},   which consists of a mostly left-handed stop, three almost degenerate Higgsino-like NLSPs ($\chi_2^0$, $\chi_3^0$ and $\chi_1^{\pm}$), and a Bino-like LSP ($\chi_1^0$).   $\tilde A_t$ is   chosen such that the SM-like Higgs mass is in the range of 125 $\sim$ 126 GeV. Even though  $\tilde A_t$ is large, the mixing between $\tilde t_L$ and $\tilde t_R$ is still small because of the large mass difference between those two components.   
 If there is a significant left-right mixing, then the $\tilde t_1 \to t \chi_2^0$ channel is highly suppressed, while the $\tilde t_1 \to \chi_1^\pm b$ channel will have a comparable branching fraction with $\tilde t_1 \to t \chi_3^0$.

\begin{table}[ht]
\begin{tabular}{|c|c|c|c|c|c|c||c|c|c|c|c|c|} \hline
    $M_1$ & $\mu$ & $M_2$ & $\tilde{A}_t$ & $M_{3SQ}$ & $M_{3SU}$ &$\tan\beta$&${\chi}_1^0$ & ${\chi}_2^0$ & ${\chi}_3^0$ & ${\chi}_1^\pm$ & $\tilde{t}_1$ & $h$ \\ \hline
     150  & 300 & 2000 & 2890 &  650  & 2000 & 10&  145 & 308 & 311 & 305 & 620 & 125 \\ \hline
\end{tabular}

\caption{Mass parameters and mass spectrum of SUSY particles for one benchmark point.  All masses are in units of GeV.  }
\label{table:MassParameters}
\end{table}

\begin{table}[h]
\centering
       \begin{tabular}{|c|c|} \hline
    Decay channel & Branching fraction  \\ \hline
    $\tilde{t}_1 \to t {\chi}_1^0$  & 3\%   \\ \hline
    $\tilde{t}_1 \to t {\chi}_2^0$  & 44\%   \\ \hline
    $\tilde{t}_1 \to t {\chi}_3^0$  & 49\%   \\ \hline
    $\tilde{t}_1 \to b {\chi}_1^+$  & 4\%   \\ \hline
   \end{tabular}
   \begin{tabular}{|c|c|} \hline
    Decay channel & Branching fraction  \\ \hline
    ${\chi}_2^0 \to Z {\chi}_1^0 $ &  96\% \\ \hline
    ${\chi}_2^0 \to h {\chi}_1^0 $ & 4\% \\ \hline
    ${\chi}_3^0 \to Z {\chi}_1^0 $ &  16\% \\ \hline
    ${\chi}_3^0 \to h {\chi}_1^0 $ & 84\% \\ \hline

   \end{tabular}
   \caption[Branching fractions for the benchmark points]{The decay branching fractions  of $\tilde{t}_1$, ${\chi}_2^0$ and ${\chi}_3^0$ for the benchmark point listed in Table~\ref{table:MassParameters}.  ${\chi}_1^\pm$  100\% decays to $W^\pm \chi_1^0$.  }
\label{table:decaychannel}
\end{table}

For this benchmark point, the decay branching fractions are shown in Table~\ref{table:decaychannel}.  $\tilde{t}_1 \to t {\chi}_{2,3}^0$  are dominant,  with branching fractions close to 50\% each, given the large ${\rm SU}(2)_L$ gauge coupling and large top Yukawa coupling of a mostly left-handed $\tilde t_1$.   The decay channels of $\tilde{t}_1 \to t {\chi}_1^0$ and $\tilde t_1 \to b {\chi}_1^+$ are highly suppressed due to the relatively small ${\rm U}(1)_Y$ gauge coupling and bottom Yukawa coupling, with branching fractions of only $3-4$\%, leading to large relaxation of the current search limits. Neutralinos ${\chi}_2^0/{\chi}_3^0$ subsequently decay to a Higgs boson or a $Z$ boson. In the case of positive $\mu$, the $\chi_2^0$ ($\chi_3^0$) dominantly decays to $Z\chi_1^0$ ($h\chi_1^0$), and reversed for negative $\mu$ value~\cite{Han:2013kza}. Therefore, changing the sign of $\mu$ has negligible impact on the collider analysis. Given the degeneracy of $\chi_2^0$ and $\chi_3^0$, the stop dominantly decays to $th\chi_1^0$ and $tZ\chi_1^0$, with branching fractions of about 45\%, respectively.  The left-handed sbottom decay modes  of $\tilde b_1 \to b \chi_2^0/\chi_3^0$ are highly suppressed due to the small bottom Yukawa coupling, while  $\tilde b_1 \to t \chi_1^\pm$ becomes dominant with branching fraction as high as 98\%. Therefore the sbottom signal will not contaminate the stop signal.

\begin{figure}[t]
\begin{center}
\includegraphics[width = 1.0 \textwidth]{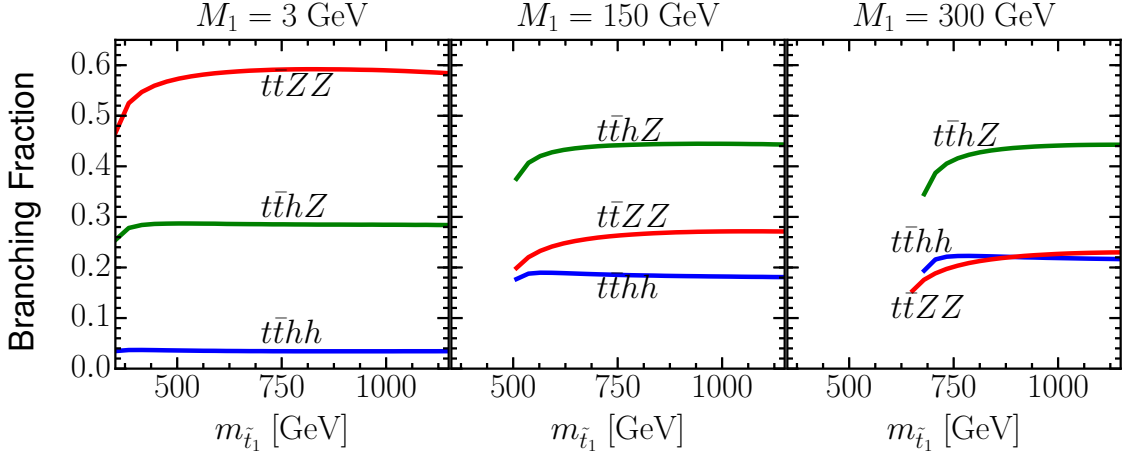}
\end{center}
\caption{The branching fractions of three different channels $t\bar t hh\met$, $t\bar t hZ\met$ and $t\bar t ZZ\met$ as a function of light stop mass for the mostly left-handed stop pair production. Three choices of $M_1$ = 3, 150, 300 GeV are presented, with $\mu$ fixed to be $M_1$ + 150 GeV and $\tan\beta=10$.}
\label{Figure:BR}
\end{figure}

At the LHC, the $\tilde t_1 \tilde t_1^*$ pair production leads to interesting final states of $t\bar{t}hh \met$, $t\bar{t}hZ \met$ and $t\bar{t}ZZ\met$.  The branching fractions are shown in  Fig.~\ref{Figure:BR}   for three different $M_1$ values of 3, 150 and 300 GeV, with $\mu$ = 150 GeV + $M_1$. When $M_1$ is small,  $\chi_{2,3}^0$ decay more to $Z\chi_1^0$,    consequently leading to a suppressed channel $t \bar t hh\met$, as shown in the left panel of Fig.~\ref{Figure:BR}.    As $M_1$ increases,   $BR(\chi_2^0/\chi_3^0 \to h \chi_1^0$) $\approx$ $BR(\chi_2^0/\chi_3^0 \to Z \chi_1^0$) $\approx$ 50\%.  The branching fraction of $\tilde t_1 \to t \chi_2^0/\chi_3^0 \to t h \chi_1^0$ and $\tilde t_1 \to t \chi_2^0/\chi_3^0 \to t Z \chi_1^0$ are almost equal, about 45\% each.     The resulting branching fractions for $t\bar{t}hh\met$ and $t\bar{t}ZZ\met$ are about 20\%, respectively, while $BR(t\bar{t}hZ\met)$ is about 45\%.

\section{Current  collider search limits on stop and recasting experimental results}
\label{sec:limit}

\subsection{Current  collider search limits on stop}

Searches for direct stop pair production have been performed at both ATLAS and CMS, with the latest results using about 36 ${\rm fb}^{-1}$ data at $\sqrt{s}=13$ TeV~\cite{Aaboud:2017aeu,Aaboud:2017ayj,Aaboud:2017wqg,Aaboud:2017nfd,Aaboud:2017dmy,Sirunyan:2017xse,Sirunyan:2017leh,Sirunyan:2017cwe,Sirunyan:2017kqq,Sirunyan:2017pjw,CMS-PAS-SUS-16-052, Sirunyan:2018iwl,Sirunyan:2017kiw,Sirunyan:2017wif}.  We summarize the current search bounds in Table \ref{tab:mbounds}.   

The current searches for
the stop mainly focus on the decay channel $\tilde t_1 \to t \chi_1^0$ and $\tilde t_1 \to b \chi_1^\pm \to b W^{(*)} \chi_1^0$,   assuming a 100\% decay branching fraction into these two channels.  Hadronic, semileptonic, and dileptonic channels have been analyzed, with the semileptonic channel typically providing the best limit.  The upper limits on the stop mass are about 1120 GeV, depending on the assumption of the decay branching fractions, and masses of the neutralinos and charginos.

In addition to the above two searching channels, the ATLAS and CMS groups also used two different analysis strategies to optimize the search sensitivity of direct stop searches for the decay channels of $\tilde t_1 \to c \chi_1^0$ and $\tilde t_1 \to b f f^\prime \chi_1^0$ , in particular, for small mass splitting between stop and $\chi_1^0$.    The upper limit on the stop mass is much weaker, about 580  GeV at 95\% C.L.~\cite{CMS-PAS-SUS-16-052, Sirunyan:2018iwl,Sirunyan:2017kiw,Sirunyan:2017wif,Aaboud:2017aeu}.

\begin{table}[tb]
\begin{tabular}{|c|c|c|c|c|}
\hline
&\multicolumn{3}{c|}{$\tilde{t}_1\rightarrow t \chi_1^0$, $\tilde{t}_1\rightarrow b \chi_1^\pm \rightarrow bW\chi_1^0$}&$\tilde{t}_1\rightarrow c \chi_1^0$, $\tilde{t}_1\rightarrow b f f^\prime \chi_1^0$ \\ \cline{2-4}
&0 $\ell$&1 $\ell$&2 $\ell$ & \\ \hline
ATALS& 1000 GeV& 940 GeV& 720 GeV & 400 GeV \\  \hline
 CMS&1070 GeV & 1120 GeV & 800 GeV & 580 GeV \\ \hline
\end{tabular}
\caption{Current mass bounds on the stop (with a small $m_{\chi_1^0}$)  from the direct searches at the 13 TeV LHC  with 36 ${\rm fb}^{-1}$ integrated luminosity~\cite{Aaboud:2017aeu,Aaboud:2017ayj,Aaboud:2017wqg,Aaboud:2017nfd,Aaboud:2017dmy,CMS-PAS-SUS-16-052,Sirunyan:2017wif,Sirunyan:2017xse,Sirunyan:2017leh,Sirunyan:2017kiw,Sirunyan:2017cwe,Sirunyan:2017kqq,Sirunyan:2017pjw,Sirunyan:2018iwl}. 
The 0$\ell$, 1$\ell$ and 2$\ell$ mean the all-hadronic, semileptonic and dileptonic final states.
}
\label{tab:mbounds}
\end{table}

\subsection{Recasting CMS search results}

Both ATLAS and CMS groups performed the search for the heavier stop ($\tilde t_2$)~\cite{Aaboud:2017ejf,Khachatryan:2014doa}  with cascade decays of $\tilde t_2 \to \tilde t_1 h$ and/or  $\tilde t_2 \to \tilde t_1 Z$ with $\tilde t_1$ further decaying via $\tilde t_1 \to t \chi_1^0$ assuming mass relation $m_{\tilde t_1} - m_{\chi_1^0} = m_t$, leading to the finals states of $t\bar{t}hh\met$, $t\bar{t}hZ\met$ and $t\bar{t}ZZ\met$ for the pair production of $\tilde t_2$ at the LHC. The analysis of the CMS group is based on the multiplicities of the leptons,  jets,  $b$-jets, missing energy $\met$, transverse mass $m_T$ and $H_T$, as demonstrated in Table I in Ref.~\cite{Khachatryan:2014doa}.  The signal regions included in their analysis are: one charged lepton (1$\ell$), two opposite-sign charged leptons (2 OS $\ell$), two same-sign charged leptons (2 SS $\ell$) and at least three charged leptons ($\ge 3 \ell$). The at least three leptons signal region is further split into two signal regions: on-$Z$, when there is a pair of same flavor, opposite-sign  charge leptons that has an invariant mass within 15 GeV of the nominal $Z$ boson mass; and off-$Z$, where no such lepton pair exists or the invariant mass lies outside the $Z$ mass window. The background predictions and observed data yields for signal regions are listed in Table II, III, IV in Ref.~\cite{Khachatryan:2014doa}.

\begin{figure}[!htbp]
\begin{center}
\includegraphics[width = 0.48 \textwidth]{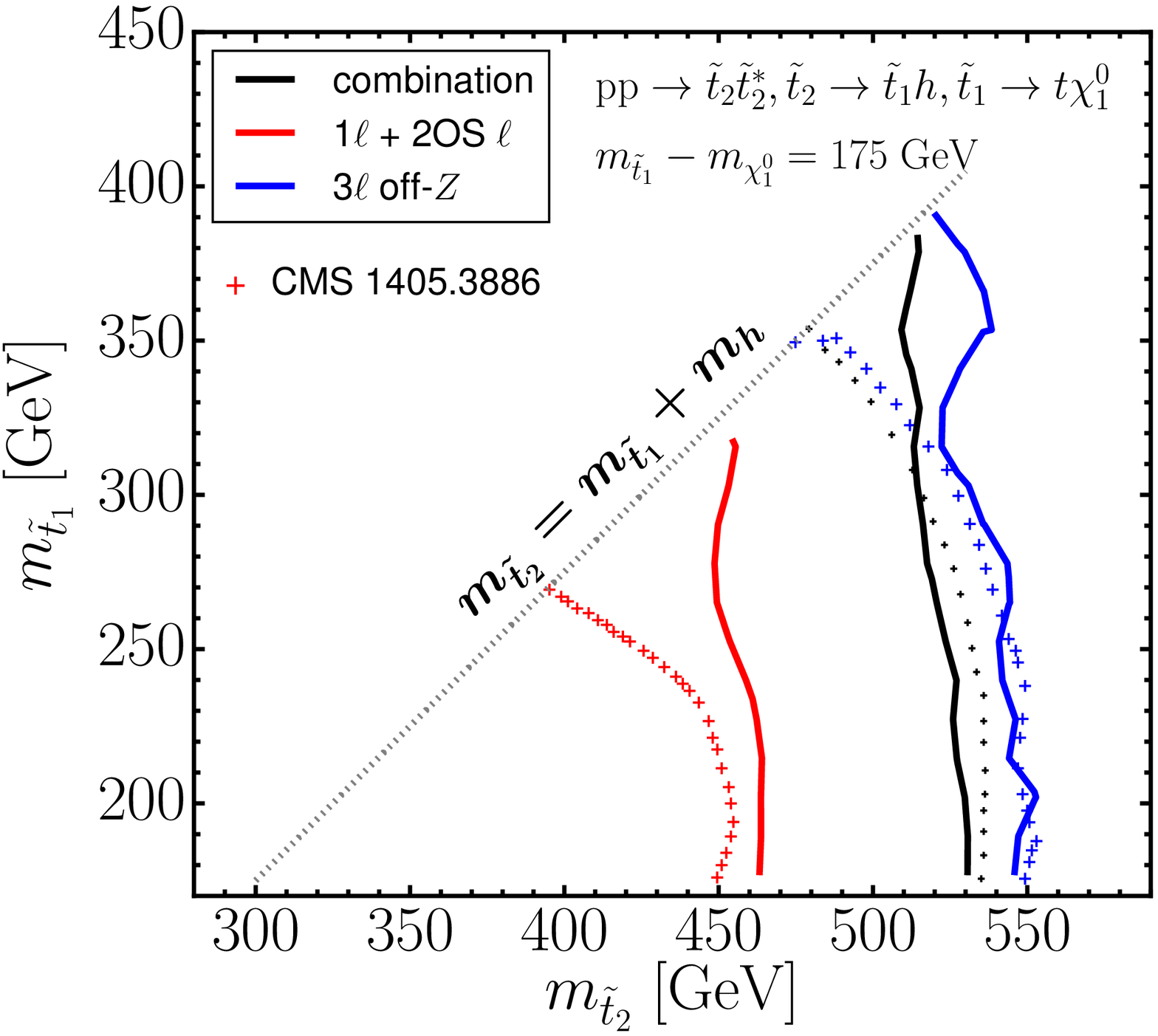}
\includegraphics[width = 0.48 \textwidth]{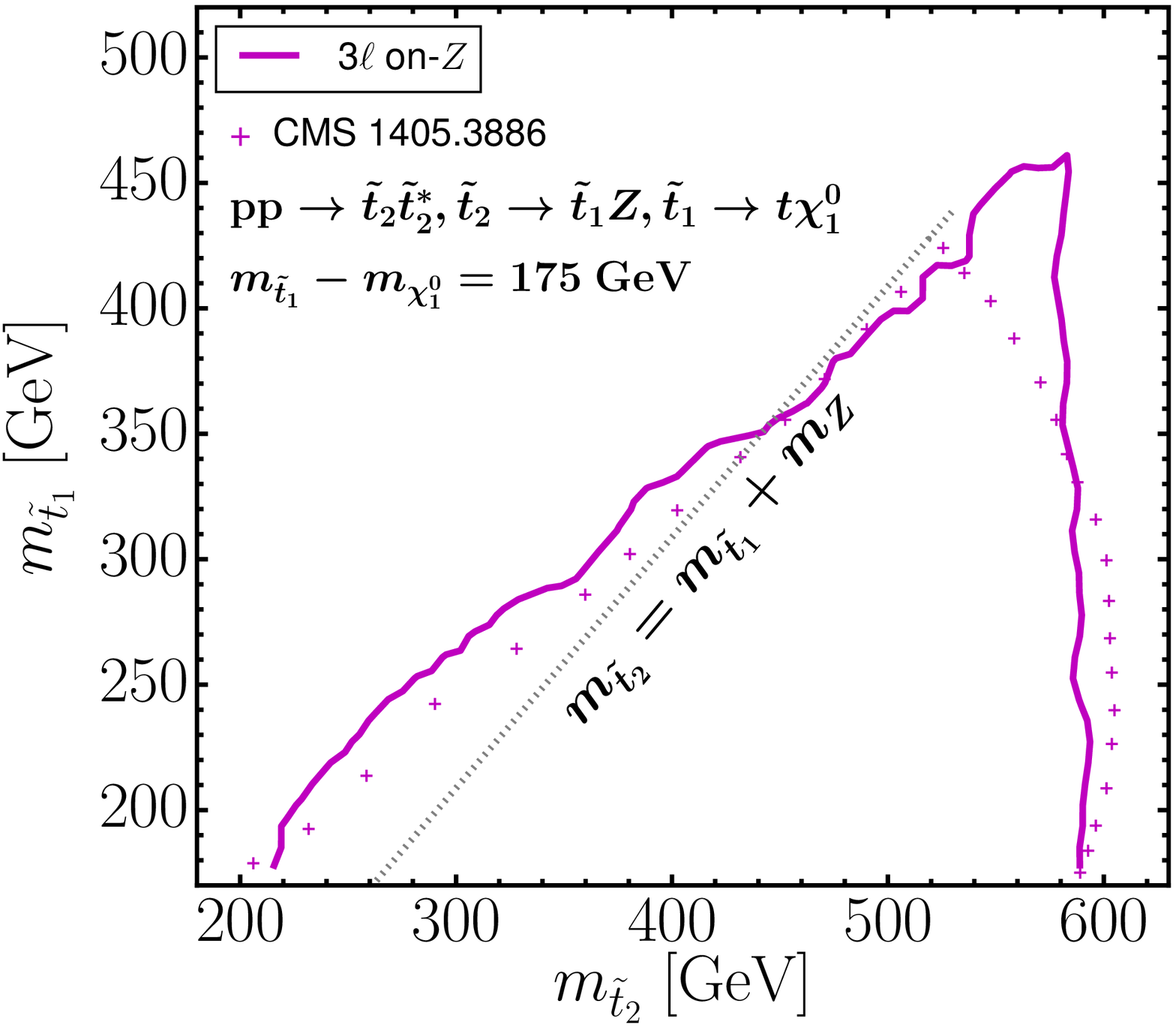}
\end{center}
\caption{The comparison of 95\% C.L. upper limits between CMS results (``+" symbol lines) and our simulations (solid lines) for the LHC  $\tilde{t}_2$ pair production, with $\tilde{t}_2\rightarrow \tilde{t}_1 h/Z$ and $\tilde{t}_1\rightarrow t \chi_1^0$.  The LSP mass is  fixed to be $m_{\tilde t_1} - m_{\chi^0_1}$ = 175 GeV.  ${\rm BR}(\tilde t_2 \to \tilde t_1 h)=100\%$ is assumed for the left panel and ${\rm BR}(\tilde t_2 \to \tilde t_1 Z)=100\%$ is assumed for the right panel.   Results~\cite{Khachatryan:2014doa} from the 8 TeV LHC with 19.5  ${\rm fb}^{-1}$ are used here.    }
\label{Figure:CMS_ttHH}
\end{figure}

We first reproduce the CMS exclusion limits for $\tilde{t}_2$ as a validation of our analyses.   Event samples are generated using Madgraph 5 \cite{Alwall:2014hca}, processed through
Pythia 6 \cite{oai:arXiv.org:hep-ph/0603175} for the fragmentation and hadronization and then through Delphes 3 \cite{deFavereau:2013fsa} for the detector simulation.    The root package TLimit \cite{Junk:1999kv} is used to calculate the 95\% confidence level upper limits.      Fig.~\ref{Figure:CMS_ttHH} shows the comparison of the 95\% C.L. upper limits between CMS results (``+" symbol lines)  \cite{Khachatryan:2014doa} and our simulations (solid lines)  in the plane of $m_{\tilde t_2}$ vs. $m_{\tilde t_1}$ for $\tilde t_2 \to \tilde t_1 h$ (left) and $\tilde t_2 \to \tilde t_1 Z$ (right) assuming a 100\% branching fraction.      Our simulations match the CMS results quite well except for the edge region. The discrepancy between the CMS results and our simulations are mostly  due to the different detector simulations of the signal process and systematics estimation. 

\begin{figure}[h]
\begin{center}
\includegraphics[width = 0.48 \textwidth]{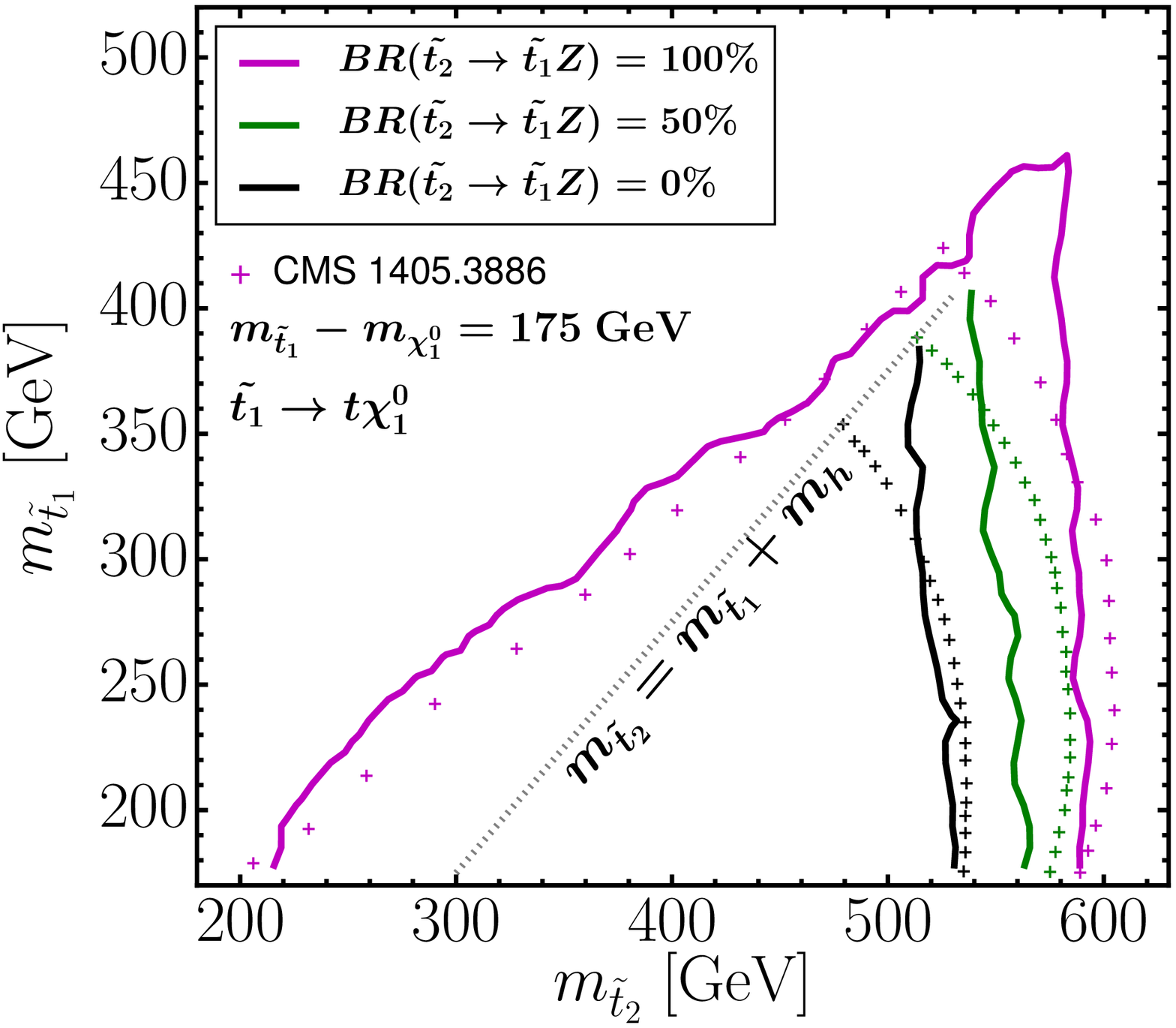}
\includegraphics[width = 0.48 \textwidth]{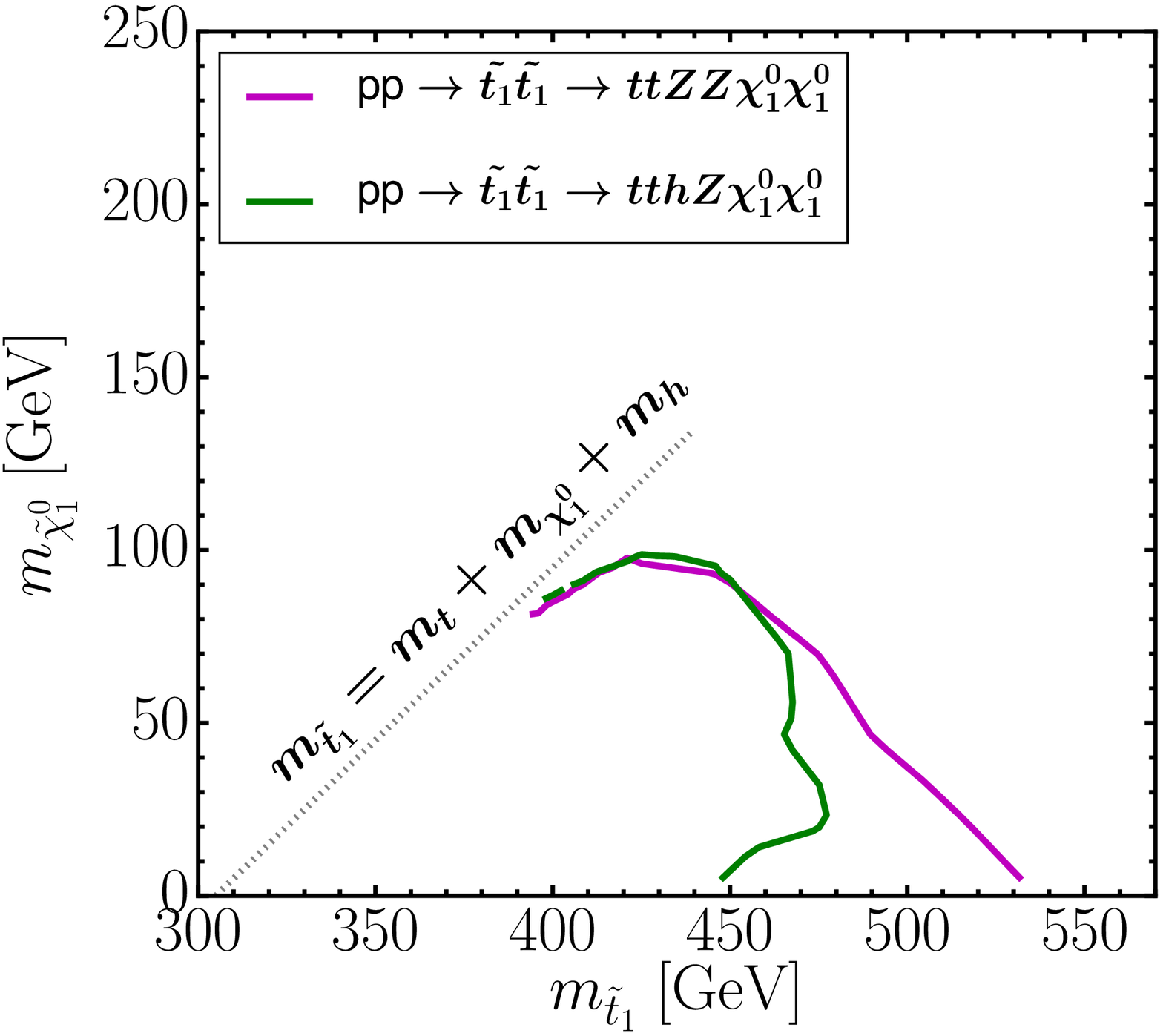}
\end{center}
	\caption{The comparison of 95\% C.L. upper limits between CMS results (``+" symbol lines) and our simulations (solid lines) for the LHC  $\tilde{t}_2$ pair production, with combined $ \tilde{t}_2\rightarrow \tilde{t}_1 h/Z$ and $\tilde{t}_1\rightarrow t \chi_1^0$, assuming $BR(\tilde t_2 \to \tilde t_1 Z)$ + $BR(\tilde t_2 \to \tilde t_1 h)$ = 100\%.    Results~\cite{Khachatryan:2014doa}  from the 8 TeV LHC with 19.5  ${\rm fb}^{-1}$ are used here.  Right panel shows the recast of the CMS $\tilde{t}_2$ limits to the plane of $m_{\tilde{t}_1}$ vs. $m_{\chi_1^0}$, considering $\tilde{t}_1\tilde{t}_1^*$ production with decays of $\tilde{t}_1\rightarrow t \chi_{2,3}^0\rightarrow th/Z \chi_1^0$.}
\label{Figure:CMS_com}
\end{figure}

Combining both $h$- and $Z$-channel, the  95\% C.L. upper limits in the plane of $m_{\tilde t_2}$ vs $m_{\tilde t_1}$ for  $BR(\tilde t_2 \to \tilde t_1 Z)=$100\% (purple), 50\% (black) and 0\% (red) are shown in the left panel of Fig.~\ref{Figure:CMS_com}, for the comparison between the CMS results and our simulations.     The decay channel of $\tilde t_2 \to \tilde t_1 h$ is only considered when the Higgs boson production is kinematically open.

Since the CMS $\tilde{t}_2$ search channel has the same final states as our $\tilde{t}_1$ study: $\tilde{t}_1\tilde{t}_1^*$ pair production with $\tilde{t}_1\rightarrow t \chi_{2,3}^0\rightarrow t h/Z \chi_1^0$,  we recast the CMS $\tilde t_2$ search limits at  8 TeV LHC to that of the lighter stop  in the scenario of Higgsino-NLSP and Bino-LSP.    We  use exactly the same event selections as the CMS $\tilde{t}_2$ search  to obtain our simulated signal event yields after cuts and we use the backgrounds estimations and observed data yields in Ref. \cite{Khachatryan:2014doa} to get the lighter stop search limits.  The recasted results in the plane of $m_{\tilde{t}_1}$ vs. $m_{\chi_1^0}$ are shown in the right panel of Fig. \ref{Figure:CMS_com} for $t\bar{t}ZZ\met$ (including 3$\ell$ ``on-$Z$") and $t\bar{t}hZ\met$ (including 1$\ell$, 2 OS $\ell$, 3$\ell$ ``off-$Z$" and 3$\ell$ ``on-$Z$") channels.    Because the reach of the 2 SS $\ell$  signal region is very low,  it is not considered in this analysis. There is also no excluding reach for the channel of $tt \bar hh \met$  due to its low branching fraction as shown in Fig.~\ref{Figure:BR}.  The light stop mass up to 480 GeV is excluded at 95\% C.L.  for a small mass LSP $\sim$ 25 GeV via the $t\bar{t}hZ\met$ channel. For the $t\bar{t}ZZ\met$ channel, the light stop mass up to 530 GeV  is excluded at 95\% C.L.  for a massless LSP.

\section{Collider analyses at $\sqrt{s} =$ 14 TeV}
\label{sec:analysesof14}

In   MSSM with more than one neutralino/chargino lighter than the stop,  typically more than one decay mode for stop are present, some of which even dominate the most commonly studied channels of $\tilde t_1 \to t \chi_1^0/b \chi_1^+$.  Those extra stop decay modes weaken the current search limits using $\tilde t_1 \to t \chi_1^0/b \chi_1^+$.   Furthermore, the new decay channels offer alternative discovery potential for the stops. In our analyses, we work in the scenario of a Bino LSP with Higgsino NLSPs lighter than $\tilde{t}_1$,  assuming the mass hierarchy of $M_1 < \mu < M_{3SQ} \ll M_2$.

The benchmark point shown  in Table \ref{table:MassParameters} is  only for the illustration purpose. In the following analyses, we perform a broad scan over the mass parameter space:

$\bullet$
$M_{3SQ}$ from 400 to 1250 GeV with a step size of 25 GeV, corresponding to $m_{\tilde{t}_1}$ varying from  350 GeV to about 1260 GeV.

$\bullet$
$M_1$ is scanned from 3 GeV to 750 GeV,  in the step size of 25 GeV.

$\bullet$
$\mu$ is fixed to be $\mu$ = $M_1$ + 150 GeV.

$\bullet$
We further require $m_{\tilde{t}_1}>m_{\chi_2^0}/m_{\chi_3^0}+m_t$ such that $\tilde{t}_1\to t\chi_2^0/\chi_3^0$  is kinematically open. 

Event samples including signals and all the SM backgrounds  are generated for 14 TeV LHC, using Madgraph 5 \cite{Alwall:2014hca}, processed through
Pythia 6 \cite{oai:arXiv.org:hep-ph/0603175} for the fragmentation and hadronization,  and then through Delphes 3 \cite{deFavereau:2013fsa} with the Snowmass combined LHC No-Pile-up detector card \cite{Anderson:2013kxz} for the detector simulation. Both the
SM backgrounds and the stop pair production signal are normalized to the predicted  cross sections, calculated including higher-order QCD corrections \cite{oai:arXiv.org:1006.4771, Broggio:2013uba, Borschensky:2014cia,oai:arXiv.org:0804.2800, oai:arXiv.org:0905.0110, oai:arXiv.org:hep-ph/0211352, oai:arXiv.org:1204.5678, oai:arXiv.org:0804.2220}. For the event generation, the top quark mass $m_t$ is set to be 175 GeV, and the Higgs mass $m_h$ is set to be 125 GeV.

\subsection{Event Selection}
\label{sec:event_selection}

For the stop  pair production $\tilde{t}_1\tilde{t}_1^*$ at the LHC, both stops decay via $t\chi_2^0/\chi_3^0$ with neutralinos subsequent decaying to a $Z$ boson or a Higgs boson, leading to final states of $t\bar{t}hh\met$, $t\bar{t}ZZ\met$ and $t\bar{t}hZ\met$.
Similar to the CMS $\tilde{t}_2$ searches, we divide the signal regions into three primary categories: (1) one charged lepton (1$\ell$), (2) two opposite-sign charged leptons (2OS $\ell$), (3) at least three charged  leptons  ($\ge3\ell$).  ``on-$Z$'' region and ``off-$Z$'' region are further defined for  the $\ge3\ell$ case, with $m_{\ell\ell}$ window of $m_Z\pm 15$ GeV.  The signal region of two same-sign leptons is not considered in this analysis because the cross section of this signal region is quite small, which results in limited reach of this signal region.

The jets are reconstructed using anti-$k_t$ algorithm with cone radius of 0.5. All jets are required to meet the basic selection cuts of $p_T^j >$ 30 GeV and $\eta^j <$ 2.5.  All leptons ($e$ or $\mu$) are required to meet the basic selection cuts of $\eta^\ell <$ 2.5 and $p_T^\ell >$ 10 GeV.   In addition to the selection cuts mentioned above, we also apply some advanced  cuts which are defined below:

\begin{itemize}
\item{$\met$,  the magnitude of the missing transpose momentum ${\bf p}_T^{miss}$.}
\item{$H_T$,  the scalar sum of the $p_T$ of all the jets which meet the basic selection cuts: $H_T = \sum_{jet}p_T^{j}$.}
\item $m_T$,   the invariant mass of the lepton and the missing transpose momentum:
\begin{equation}
m_T = \sqrt{2  p_T^\ell\met (1-\cos\phi(\mathbf{p}_T^\ell,\mathbf{p}^{miss}_T))}.
\end{equation}
\item{$M_{T2}$ \cite{mt2_1,mt2_2,Cheng:2008hk},  the lower bound on  the transverse mass resulting from two missing energies.
\begin{equation}
M_{T2}({\bf p}_T^{\ell_1},{\bf p}_T^{\ell_2},{\bf p}_T^{\rm miss})=\min\limits_{{\bf p}_{T,1}^{\rm miss}+{\bf p}_{T,2}^{\rm miss}={\bf p}_T^{\rm miss}}\{\max\{m_T({\bf p}_T^{\ell_1},{\bf p}_{T,1}^{\rm miss}),m_T({\bf p}_T^{\ell_2},{\bf p}_{T,2}^{\rm miss})\}\}.
\end{equation}
}
\item{$m_{\ell\ell}$, the invariant mass of two OS leptons which survive  the basic selection cuts.}
\item{$N_j$, the number of all the jets which meet the basic selection cuts.}
\item{$N_b$, the number of all the $b$ jets which meet the basic selection cuts.}
\end{itemize}

We summarize the cuts we used in Table \ref{tab:cuts}.

\begin{table}[tb]
\resizebox{\textwidth}{!}{
\hspace*{-1cm}
\begin{tabular}{|c|c|c|c|}
\toprule[1pt]
 & 1$\ell$ & 2OS $\ell$ & $\ge3\ell$ \\
\midrule[1pt]
\multirow{6}{*}{Basic cuts} & Leading three jets $p_T >$ 40  & Leading two jets $p_T >$ 40 & - \\
	& $N_j \ge 4$, $N_b \ge 2$ & $N_j \ge 4$, $N_b \ge 2$ & $N_j \ge 2$, $N_b \ge 1$ \\
	& Exact one lepton with $p_T >$ 25 & Exact two leptons with $p_T >$ 25 & $\ge3$ leptons with $p_T >$ 10 \\
	& $\Delta R(j,l) > 0.4$ & $\Delta R(j,l) > 0.4$, $\Delta R(l,l) > 0.4$ & $\Delta R(j,l) > 0.4$, $\Delta R(l,l) > 0.4$ \\
	& $\Delta \Phi(j,{\bf p}_T^{miss}) >$  0.8 & $\Delta \Phi(j,{\bf p}_T^{miss}) >$  0.8 & $\Delta \Phi(j,{\bf p}_T^{miss}) >$  0.8 \\
	& - & - & ``off-$Z$", ``on-$Z$" \\
\hline

\multirow{6}{*}{Advanced cuts} & $\met>$  100, 125, 150, 175, 200 & $\met>$  150, 175, 200, 225, 250 &  $\met>$  150, 175, 200, 225, 250  \\
	& $H_T>$  400, 450, 500, 550, 600 &  $H_T>$  400, 450, 500, 550, 600 &  $H_T>$  400, 450, 500, 550, 600 \\
	& $m_T>$  100, 125, 150, 175, 200 & $M_{T2} >$ 60, 70, 80, 90 & - \\
	& - & $|m_{\ell\ell} - m_Z| <$ 5, 10, 15 &  $|m_{\ell\ell} - m_Z| <$ 5, 10, 15 \\
	& $N_j \ge$ 4, 5, 6, 7 & $N_j \ge$ 4, 5, 6 & $N_j\ge$ 2, 3, 4, 5\\
	& $N_b\ge$ 2, 3, 4, 5 & $N_b\ge$ 2, 3, 4 & $N_b\ge$ 1, 2, 3 \\

\bottomrule[1pt]
\end{tabular}
}
	\caption[Basic cuts]{ The basic cuts and the advanced cuts for the three primary signal regions of $1\ell$, 2 OS $\ell$ and $\ge3\ell$.  All mass units are in GeV. }
\label{tab:cuts}
\end{table}

\subsection{Results of one lepton signal region}

In this section and the following sections, we focus on the discovery/exclusion reach of the light stop at the 14 TeV LHC with integrated luminosity of 300 ${\rm fb^{-1}}$. In the $ 1 \ell$  signal region, the advanced selection cuts of $\met$, $H_T$, $m_T$, $N_j$ and $N_{b}$ are used to cut down the huge SM backgrounds. Table~\ref{tab:onelep} shows the cumulative cut efficiencies after each level of advanced cuts and final cross sections for both signal as well as the SM backgrounds, for the benchmark point listed in Table.~\ref{table:MassParameters}.    As expected, the signal process has larger $m_T$ and $\met$ than the background processes due to the extra contributions from the LSP.  $t\bar{t}$, $t\bar{t}b\bar{b}$ and $t\bar{t}Z$ are the dominant backgrounds after strong $\met$, $H_T$ and  $m_T$ cuts. The irreducible SM backgrounds $t \bar t h h$, $t \bar t h Z$ and $t \bar t ZZ$ are almost negligible because of the very low production cross sections.
\begin{table}[tb]
\resizebox{\textwidth}{!}{
\begin{tabular}{|c|c|c|c|c|c|c|c|c|} 
\toprule[1pt]
Process & $\sigma$ (fb)& Basic &$\met>$ & $H_T>$ & $m_T>$ & $N_j\geq$ & $N_b\geq$ & $\sigma$ (fb)\\
        &  & cuts & $175\text{ GeV}$ & $500\text{ GeV}$ & $150\text{ GeV}$ & $7$ & $2$ & after cuts \\
\toprule[1pt]
$\tilde{t}_1 \tilde{t}_1 (t\bar t hh)$ & 35 & 11.0\% & 4.6\% & 3.6\% & 1.6\% & 0.5\% & 0.5\% & 0.175 \\
\hline
$\tilde{t}_1 \tilde{t}_1 (t\bar t hZ)$ & 80 & 8.7\% & 4.3\% & 3.4\% & 1.6\% & 0.45\% & 0.45\% & 0.36 \\
\hline
$\tilde{t}_1 \tilde{t}_1 (t\bar t ZZ)$ & 46 & 5.7\% & 3.1\% & 2.3\% & 1.3\% & 0.31\% & 0.31\% & 0.14 \\
\hline
$t\bar{t}_{\rm semi}$ & 261230 &  1.9\% &  $5.2\times10^{-4}$ &  $1.6\times10^{-4}$ & $8.4\times10^{-7}$ & $5\times10^{-8}$ & $5\times10^{-8}$ & 0.013 \\
\hline
$t\bar{t}b\bar{b}$  & 8305 & 3.2\% & 0.17\%  & $9.3\times10^{-4}$ & $7.4\times10^{-5}$ & $6.6\times10^{-6}$ & $6.6\times10^{-6}$ & 0.055\\
\hline
$t\bar{t}Z$ & 1095 & 2.3\% & 0.23\% & 0.12\% & $2.7\times10^{-4}$ & $2.2\times10^{-5}$ & $2.2\times10^{-5}$ & 0.024\\
\hline
$t\bar{t}W^{\pm}$ & 747 & 1.8\% & 0.18\% & 0.11\% & $9.7\times10^{-5}$ & $4.8\times10^{-6} $ & $4.8\times10^{-6}$ & $3.6 \times 10^{-3}$\\
\hline
$t\bar{t}h$ & 572 & 4.6\% & 0.34\% & 0.22\% & $1.6\times10^{-4}$ & $1.4\times10^{-5}$ & $1.4\times10^{-5}$ & $8.1 \times 10^{-3}$\\
\hline
$t\bar{t}hh$ & 0.83 & 10.8\% & 1.1\% & 0.87\% & 0.012\% & $1.7\times10^{-4}$ & $1.7\times10^{-4}$ & $1.5 \times 10^{-4}$\\
\hline
$t\bar{t}hZ$ & 1.41 & 7.4\% & 1.2\% & 0.85\% & 0.022\% & $3.3\times10^{-4}$ & $3.3\times10^{-4}$ & $4.6 \times 10^{-4}$\\
\hline
$t\bar{t}ZZ$ & 1.73 & 4.1\% & 0.74\% & 0.51\% & 0.016\% & $2.0\times10^{-4}$ & $2.0\times10^{-4}$ & $3.5 \times 10^{-4}$\\
 \bottomrule[1pt]
\hline
\end{tabular}
}
	\caption{cumulative cut efficiencies  after each level of cuts and the final cross sections for the signal $\tilde{t}_1\tilde{t}_1^* \rightarrow t\bar thh \met$, $t\bar thZ\met$ and $t\bar tZZ\met$,  as well as SM backgrounds for $1 \ell$  signal region for the benchmark point listed in Table.~\ref{table:MassParameters}.     Note that the cross section for $t \bar t$ is shown for the semileptonic decay only, which is the dominant $t\bar{t}$ background. }
\label{tab:onelep}
\end{table}

\begin{figure}[!htbp]
\begin{center}
\includegraphics[width = 0.5 \textwidth]{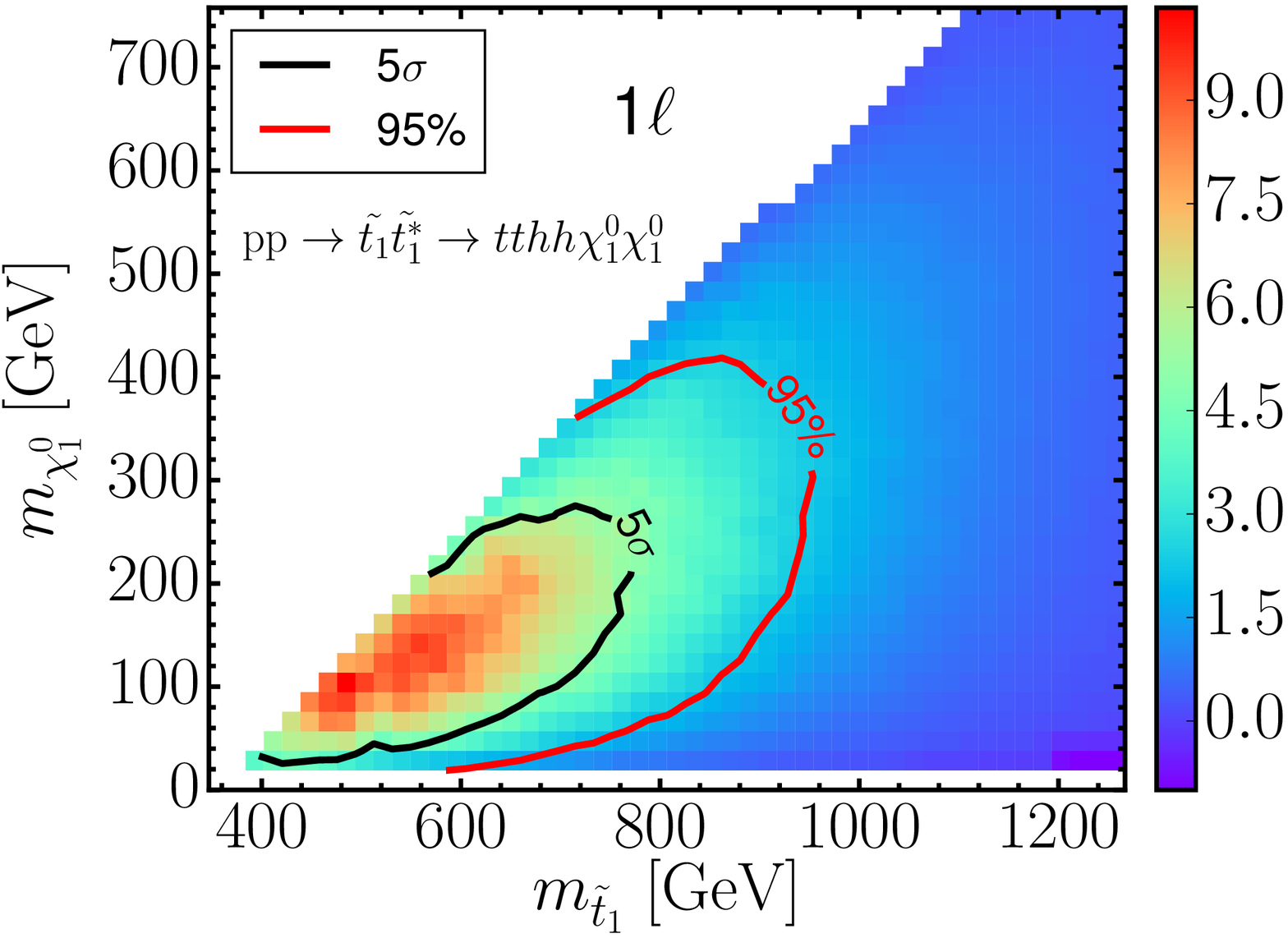}
\includegraphics[width = 0.49 \textwidth]{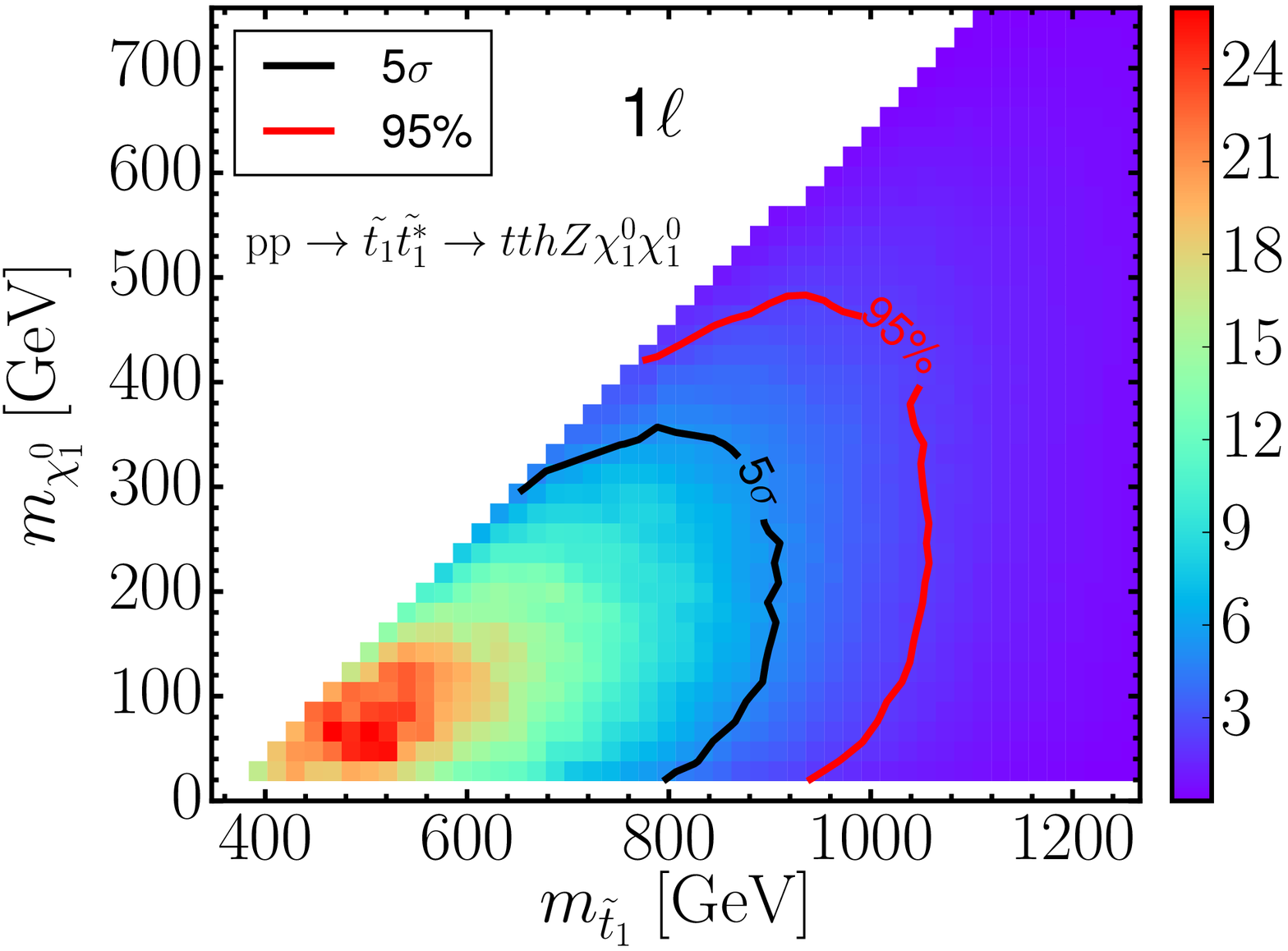}
\includegraphics[width = 0.5 \textwidth]{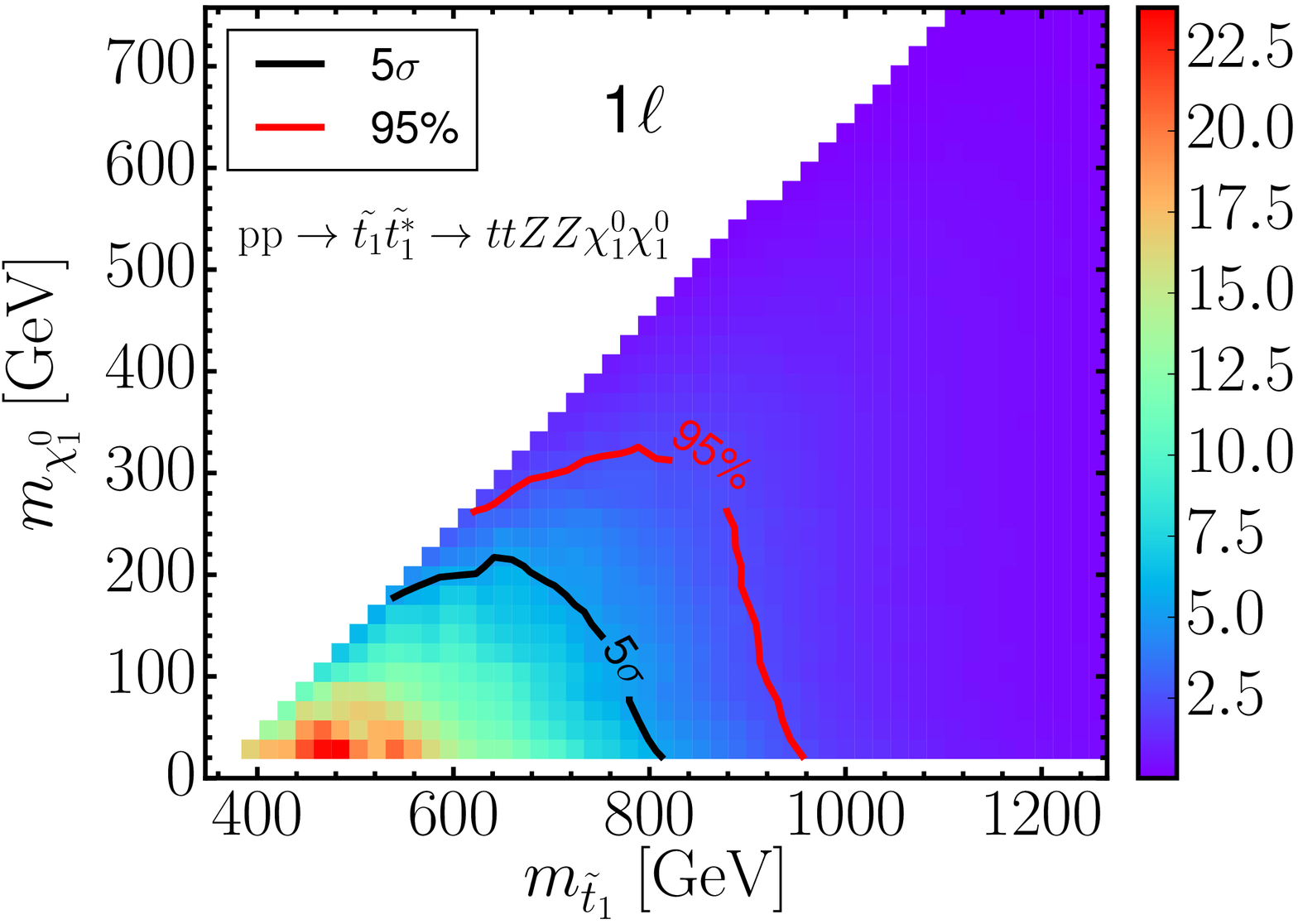}
\includegraphics[width = 0.49 \textwidth]{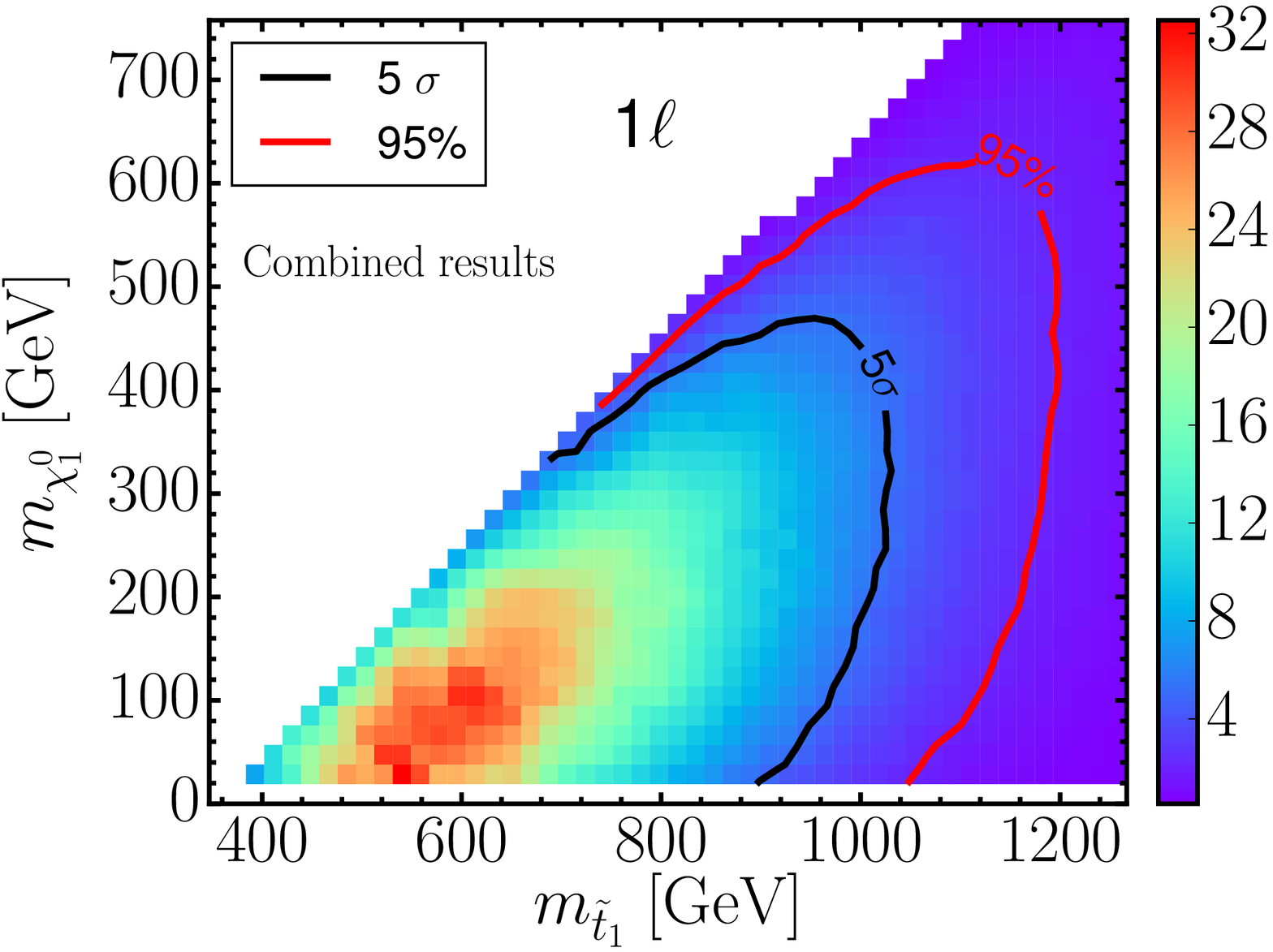}
\end{center}
	\caption[Signal significance contour for one lepton channel]{The 95\% C.L. upper limits (black) and 5$\sigma$ discovery reach (red) are shown in the plane of MSSM parameter space $m_{\tilde t_1}$ vs $m_{\chi_1^0}$ for the stop pair production $pp \to \tilde t_1 \tilde t_1^* \to t \bar t \chi_2^0/\chi_3^0 \to t \bar t h h \met$ (top left), $t \bar t hZ\met$ (top right), $t \bar t ZZ\met$ (bottom left),  combined channels (bottom right)  for the 1 $\ell$  signal region at the 14 TeV LHC with 300 ${\rm fb}^{-1}$ integrated luminosity. $\mu$ is fixed to be $M_1$ + 150 GeV. 10\% systematic error has been included in the analyses.  The color coding on the right indicates the signal significance  to guide the eye.    }
\label{Figure:reach_1lep}
\end{figure}

In Fig.~\ref{Figure:reach_1lep}, the 95\% C.L. upper limits (black curve) and 5$\sigma$ discovery (red curve) reach are shown in the plane of MSSM parameter $m_{\tilde t_1}$ vs $m_{\chi_1^0}$ for the stop pair production $pp \to \tilde t_1 \tilde t_1^* \to t \bar t \chi_2^0/\chi_3^0 \to t \bar t h h \met$ (top left), $t \bar t hZ \met$ (top right) and $t \bar t ZZ \met$ (bottom left) at the 14 TeV LHC with 300 ${\rm fb}^{-1}$ integrated luminosity. $\mu$ is fixed to be $M_1$ + 150 GeV and 10\% systematic uncertainties are assumed.       All combinations of the values of advanced  cuts for $\met$, $H_T$, $m_T$, $N_j$ and $N_{bj}$, as given in Table.~\ref{tab:cuts}, are examined.  The optimized combination that gives the best significance  is used for a particular mass point.  The channel $t \bar t hh\met$ has no sensitivity in the low $\chi_1^0$ mass region because of the very low branching fraction of the $t \bar t hh\met$ channel. In contrary, the channel $t \bar t ZZ\met$ has the largest reach in the low $\chi_1^0$ mass region due to its large branching fraction. The $t \bar t hZ\met$ has the best reach in the whole mass parameter region because of its comparably large branching fraction.  For the channel $t \bar t hh\met$, stop masses up to 750 GeV can be discovered at the 5~$\sigma$ significance level for $m_{{\chi}_1^0}$ = 220 GeV, and the 95\% C.L. exclusion limits  are about  950 GeV for $m_{{\chi}_1^0} =$ 250 GeV. The 5 $\sigma$ discovery reach can go up to 900 GeV (820 GeV), or the stop masses up to 1050 GeV (960 GeV) can be excluded at the 95\% C.L. for the channel $t \bar t hZ\met$ ($t \bar t ZZ\met$).   Limits with 20\% systematic uncertainties  are about 100 GeV worse. 

The combination of all three channels gives better reach, which is shown in the bottom right panel of Fig.~\ref{Figure:reach_1lep}.  The specific set of advanced selection cuts used to do the signal combinations are: $\met>200$ GeV, $H_T>$ 550 GeV, $m_T>200$ GeV,  $N_j \geq 7$ and $N_{bj}\geq 2$.  
  The stop mass can be discovered at 5$\sigma$ significance up to 1030 GeV, or excluded at 95\% C.L. up to 1200 GeV for the $ 1 \ell$  signal region.

\subsection{Results of 2 OS $\ell$  signal region}

\begin{figure}[!htbp]
\begin{center}
\includegraphics[width = 0.48 \textwidth]{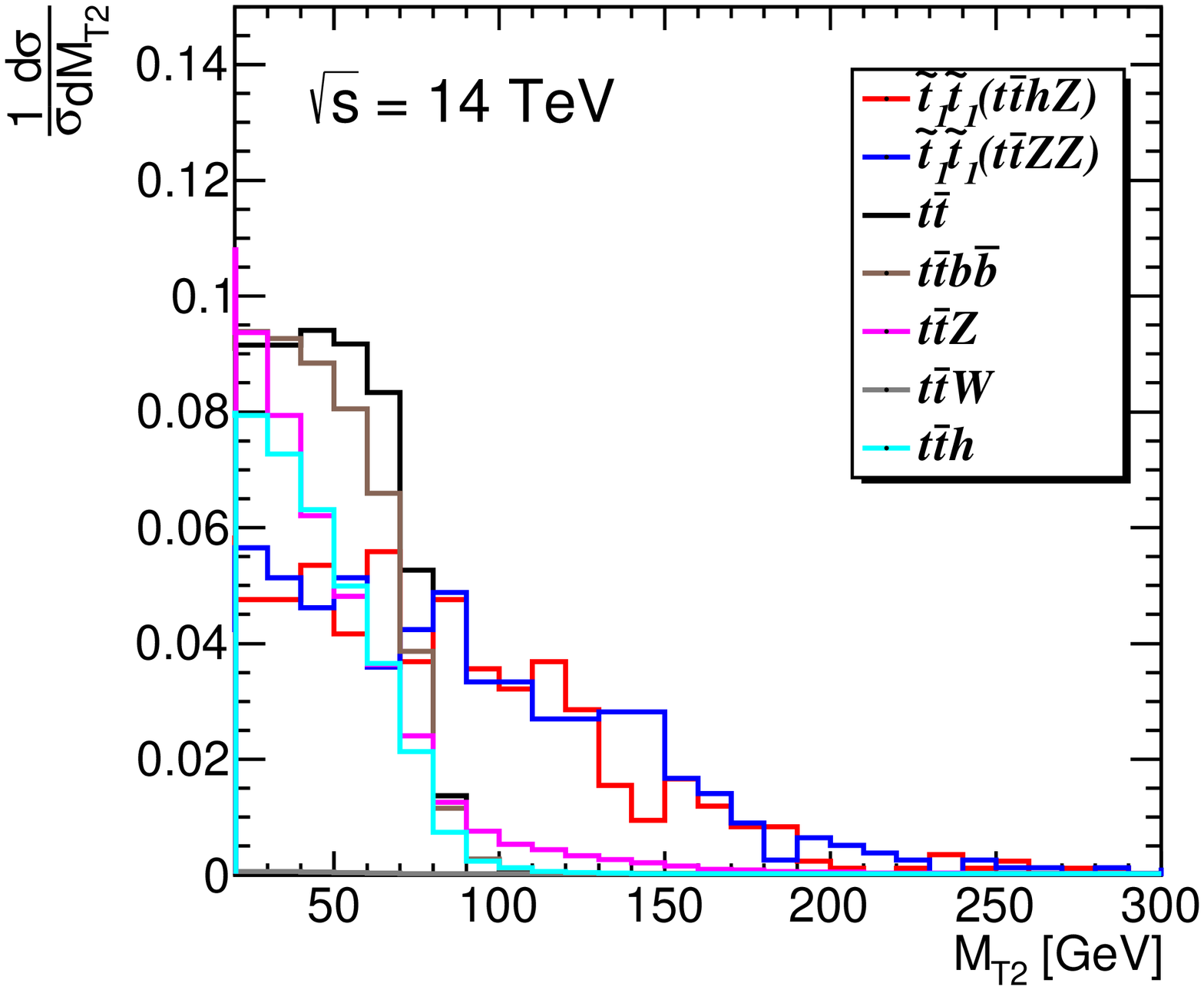}
\includegraphics[width = 0.48 \textwidth]{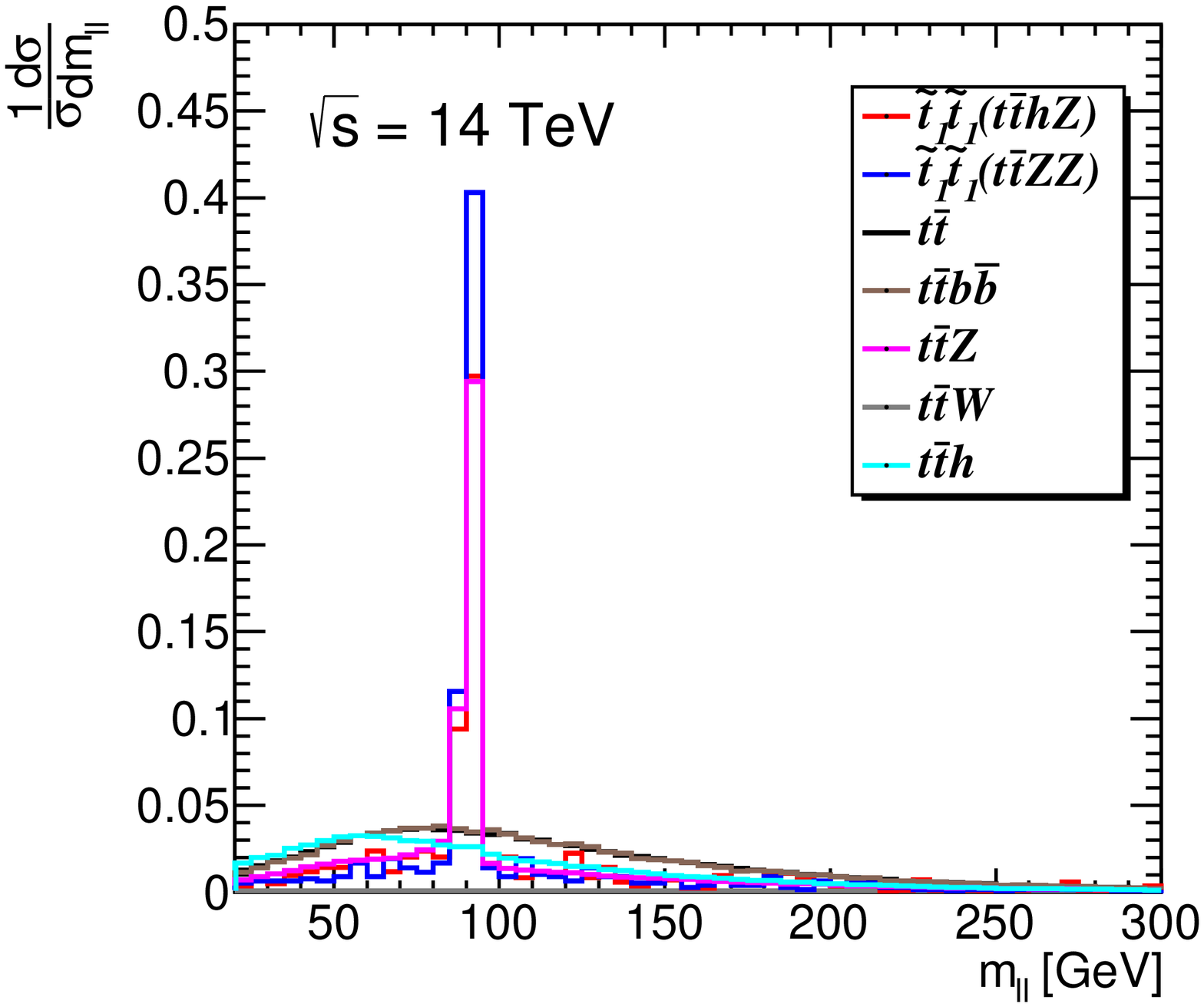}
\end{center}s\caption[Normalized distributions of $M_{T2}$ and $m_{\ell\ell}$]{Normalized distributions of $M_{T2}$ (left) and $m_{\ell\ell}$ (right) for the signal channels $t \bar t hZ$  and $t \bar t ZZ$  with $m_{\tilde t_1} =$ 620 GeV and the SM backgrounds after basic selection cuts. }
\label{Figure:plots_2lep}
\end{figure}

In the 2 OS $\ell$ signal region, in addition to the advanced  cuts of $H_T$, $\met$, $M_{T2}$, $N_j$ and $N_{bj}$, $m_{\ell\ell}$ are also used for the  $t \bar t h Z\met$ and $t \bar t ZZ\met$ channels.  
The normalized distributions of $M_{T2}$ and $m_{\ell\ell}$ for the signal processes and the SM backgrounds are shown in Fig.~\ref{Figure:plots_2lep}. The $M_{T2}$ distribution for the signal extends to larger value, while the $M_{T2}$ distributions for the SM backgrounds are cut off at $m_W$ given that the two leptons of the SM backgrounds mostly come from leptonic $W$ Decay. The $m_{\ell\ell}$ distribution for the signals and SM $t \bar t Z$ has a sharp peak at the $Z$ boson mass,  while the $m_{\ell\ell}$ distributions for the other SM backgrounds   spread out because the two leptons are not from the $Z$ boson decay.

Table~\ref{tab:twolepton} illustrates the cumulative cut efficiencies after each level of advanced cut and the final cross sections for the signals and SM backgrounds in the 2 OS $\ell$ signal region for the benchmark point. The dominant background in the 2 OS $\ell$ signal region is $t \bar t Z$, given its  relatively large cross section and similar final states to the signal processes.  $t \bar t$ is the second dominant background due to its large cross section. A significance of about 12$\sigma$ (7.7$\sigma$) can be reached for signal channel $t \bar t hZ\met$ ($t \bar t ZZ\met$) for the Table.~\ref{table:MassParameters}  benchmark point at the 14 TeV LHC with 300 ${\rm fb}^{-1}$ integrated luminosity, including 10\% systematic error.

\begin{figure}[!htbp]
\begin{center}
\includegraphics[width = 0.5 \textwidth]{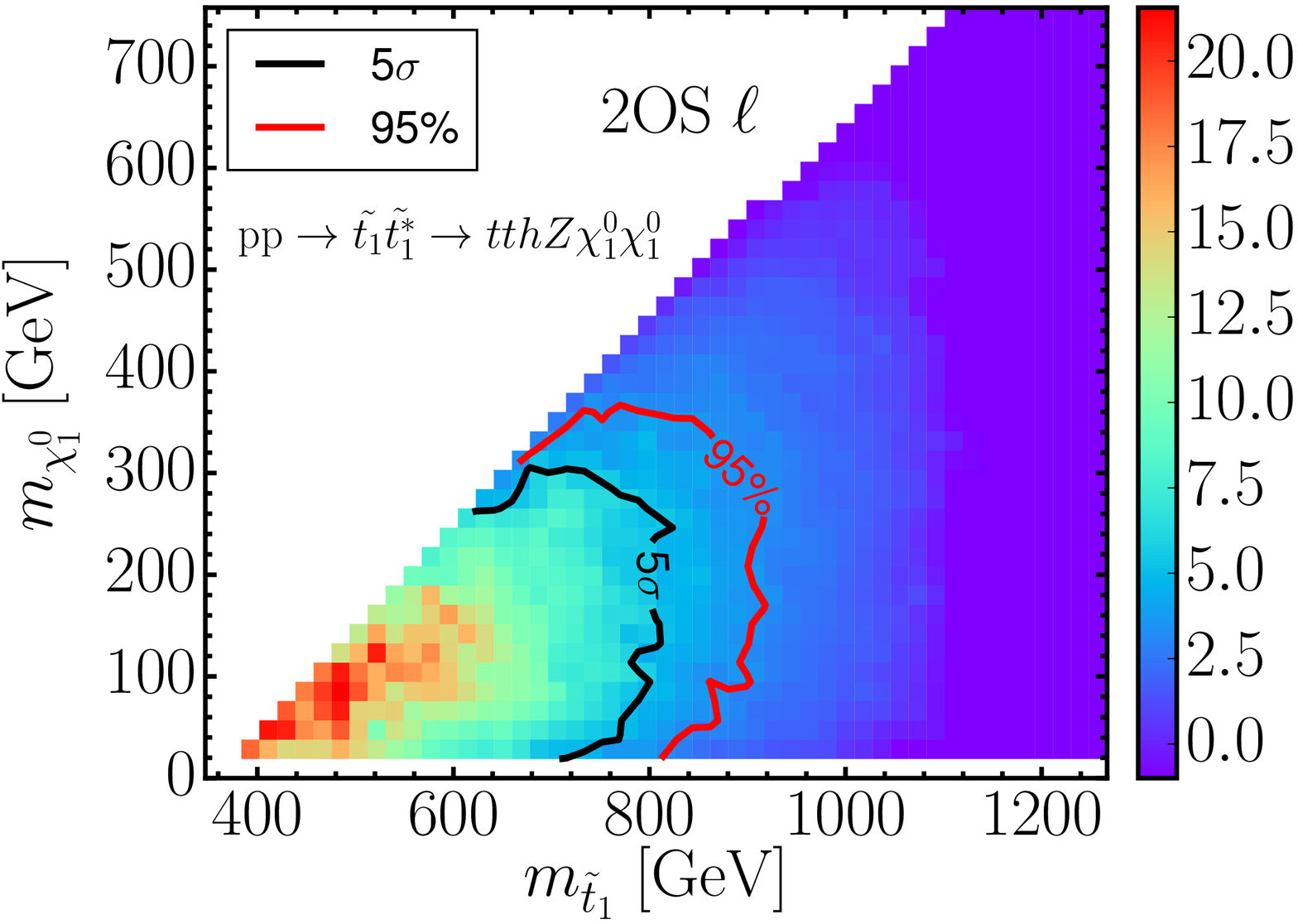}
\includegraphics[width = 0.48 \textwidth]{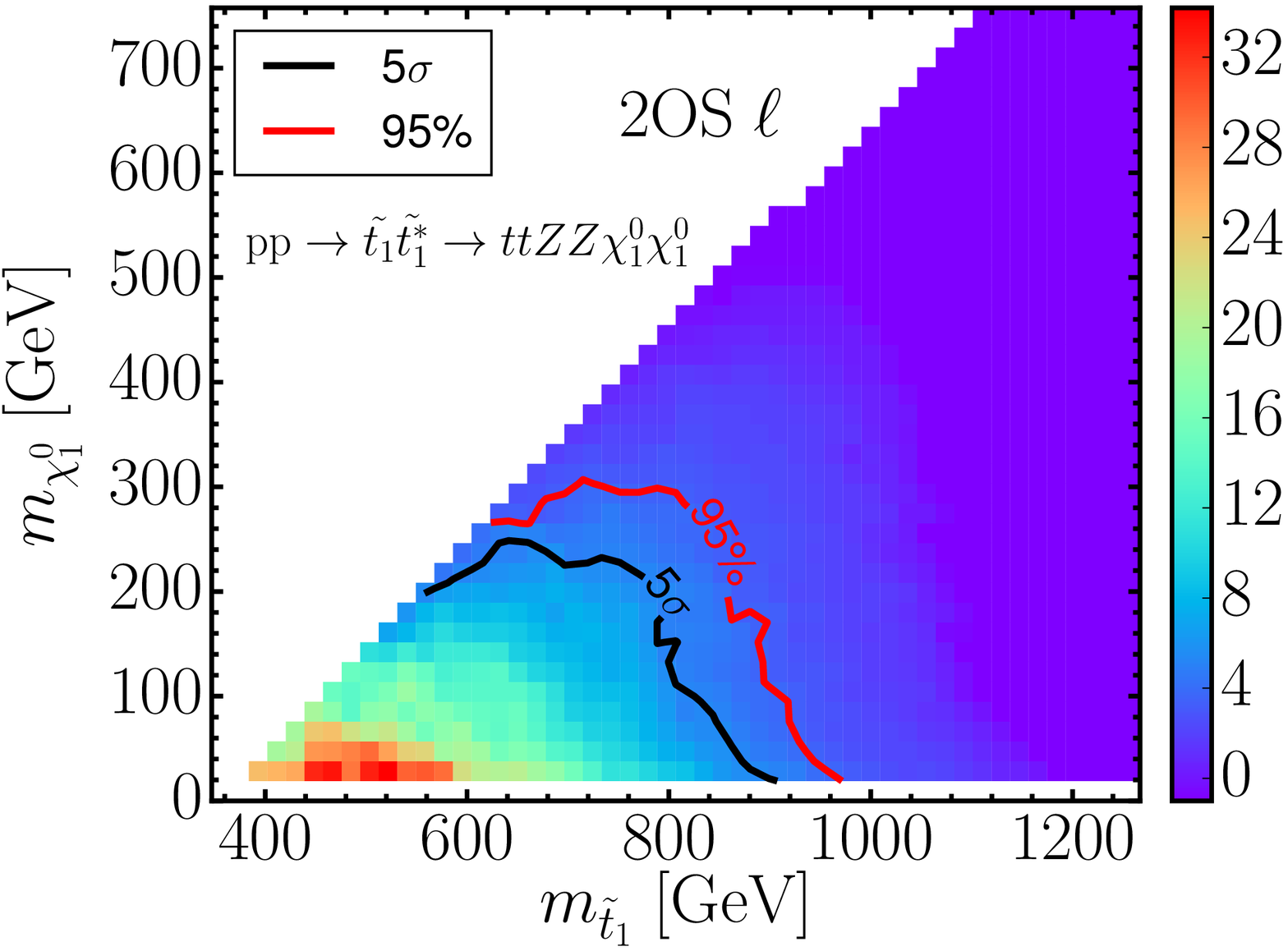}
\includegraphics[width = 0.48 \textwidth]{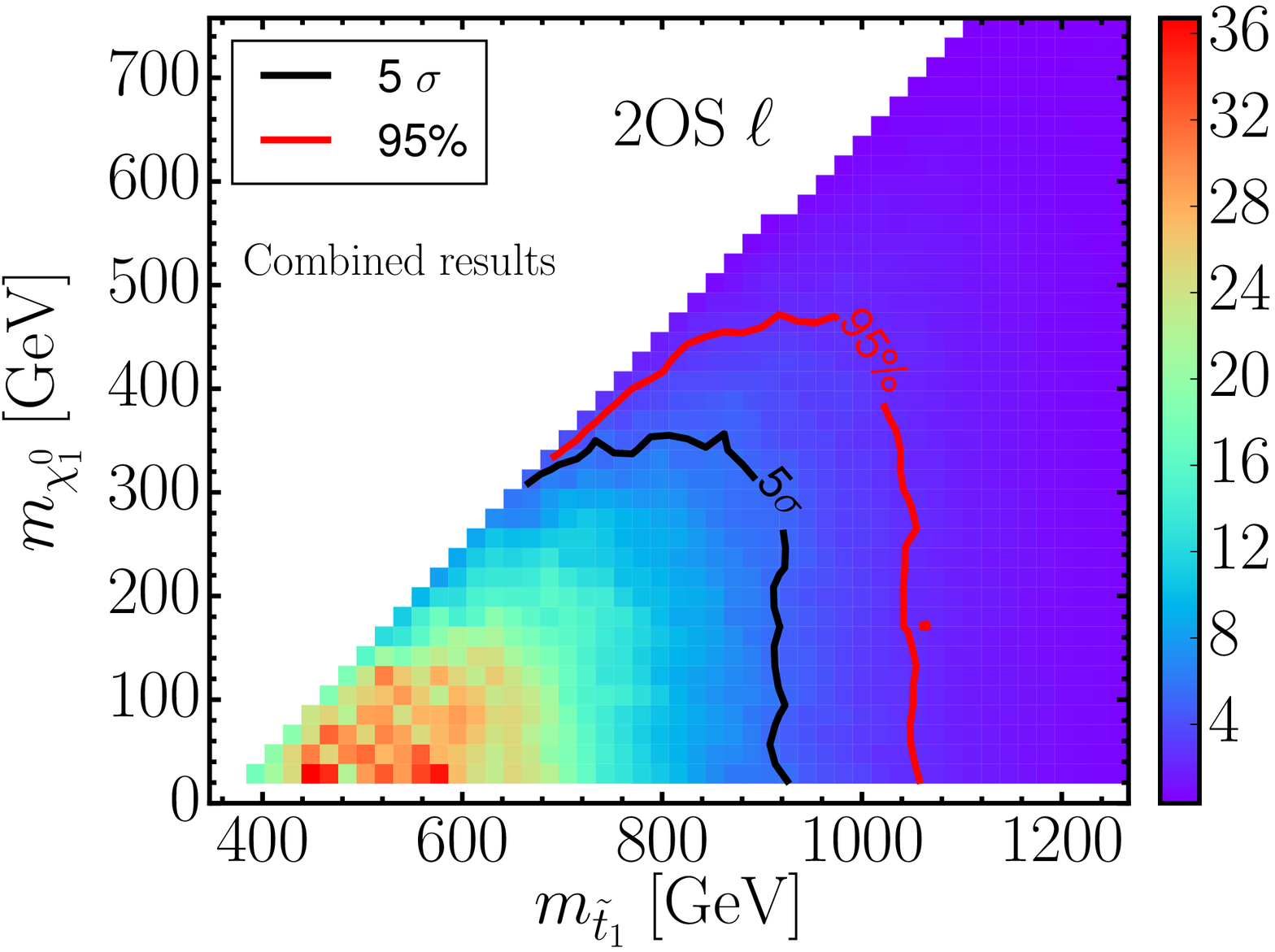}
\end{center}
\caption{The 95\% C.L. upper limits (black) and 5$\sigma$ discovery reach (red) are shown in the plane of MSSM parameter space $m_{\tilde t_1}$ vs $m_{\chi_1^0}$ for the stop pair production $pp \to \tilde t_1 \tilde t_1^* \to t \bar t \chi_2^0/\chi_3^0 \to t \bar t h Z \met$ (top left) and $t \bar t ZZ\met$ (top right) and combined reach (bottom middle) in the 2 OS $\ell$ signal region at 14 TeV LHC with 300 ${\rm fb}^{-1}$ integrated luminosity.   }
\label{Figure:reach_ttHH_2lep}
\end{figure}

For the 2 OS $\ell$ signal region, the 5$\sigma$ discovery reach (red curve) and 95\% C.L. exclusion limit (black curve) are shown in  Fig.~\ref{Figure:reach_ttHH_2lep}  for the 14 TeV LHC with 300 ${\rm fb}^{-1}$ integrated luminosity, including 10\%  systematic uncertainties.   The channel $t \bar t h h\met$ has no reach   because of its low branching fraction of the dilepton channel. A stop mass up to 800 GeV (920 GeV) can be discovered at 5$\sigma$ significance, and excluded up to  900 GeV (980 GeV) at 95\% C.L.  for the channel $t \bar t hZ\met$ ($t \bar t ZZ\met$).   Limits with 20\% systematic uncertainties are very similar to that of 10\% case since the error is mostly statistically dominated.

\begin{table}[tb]
\centering
 
\footnotesize\setlength{\tabcolsep}{2.0pt}
\begin{tabular}{|c|c|c|c|c|c|c|c|c|c|} 
\toprule[1pt]
Process & $\sigma$ (fb)& Basic &$\met>$ & $H_T>$ & $M_{T2}>$ & $|m_{\ell\ell}-m_Z| <$ & $N_j\geq$ & $N_b\geq$ & $\sigma$ (fb)\\
        &  & cuts & $100\text{ GeV}$ & $400\text{ GeV}$ & $80\text{ GeV}$ & 5 GeV & 6 & 2 & after cuts \\
\toprule[1pt]
\hline
$\tilde{t}_1 \tilde{t}_1 (t\bar t hZ)$ & 80 & 1.7\% & 1.3\% & 1.1\% & 0.34\% & 0.2\% & 0.1\% & 0.1 & 0.08 \\
\hline
$\tilde{t}_1 \tilde{t}_1 (t\bar t ZZ)$ & 46 & 1.6\% & 1.2\% & 1.1\% & 0.35\% & 0.24\% & 0.11\% & 0.11\% & 0.05 \\
\hline
$t\bar{t}_{\rm di-lep}$ & 33330 &  0.4\% &  0.14 &  $5\times10^{-4}$ & $2\times10^{-5}$ & $1\times10^{-6}$ & $9\times10^{-8}$ & $9\times10^{-8}$ & 0.003 \\
\hline
$t\bar{t}b\bar{b}$  & 8305 & 0.18\% &  $6\times10^{-4}$ & $3\times10^{-4}$ & $1\times10^{-5}$ & $4\times10^{-7}$ & $1.2\times10^{-7}$ & $1.2\times10^{-7}$ & 0.001\\
\hline
$t\bar{t}Z$ & 1095 & 0.4\% & $9\times10^{-4}$ & $5.3\times10^{-4}$ & $5\times10^{-5}$ & $2.7\times10^{-5}$ & $5.4\times10^{-6}$ & $5.4\times10^{-6}$ & 0.006\\
\hline
$t\bar{t}W^{\pm}$ & 747 & 0.2\% & $9.3\times10^{-4}$ & $5.1\times10^{-4}$ & $1.1\times10^{-5}$ & $3.2\times10^{-7}$ & $1.2\times10^{-7}$ & $1.2\times 10^{-7}$  & - \\
\hline
$t\bar{t}h$ & 572 & 0.6\% & 0.2\% & 0.1\% & $2.9\times10^{-5}$ & $1.2\times10^{-6}$ & $3.5\times10^{-7}$ & $3.5\times10^{-7}$  & - \\
\hline
$t\bar{t}hh$ & 0.83 & 3.1\% & 1.2\% & 1.0\% & $3.5\times10^{-4}$ & $1.9\times10^{-5}$ & $4.8\times10^{-6}$ & $2.5\times10^{-6}$ & -\\
\hline
$t\bar{t}hZ$ & 1.41 & 2.2\% & 0.8\% & 0.7\% & $6\times10^{-4}$ & $2.4\times10^{-4}$ & $5.9\times10^{-5}$ & $5.9\times10^{-5}$ & -\\
\hline
$t\bar{t}ZZ$ & 1.73 & 1.2\% & 0.4\% & 0.3\% & $6\times10^{-4}$ & $3\times10^{-4}$ & $9\times10^{-5}$ & $9\times10^{-5}$ & -\\
\bottomrule[1pt]
\hline
\end{tabular}
\caption{Cumulative cut efficiencies after each level of advanced selection cuts and cross sections for the signal $\tilde{t}_1\tilde{t}_1^* \rightarrow t \bar t hZ\met$ and $t \bar tZZ\met$ as well as SM backgrounds in the 2 OS $\ell$ signal region at the 14 TeV LHC.  $t \bar t hh\met$ is not listed here due to the small significance.  Note that only dileptonic decay of $t \bar t$ is in used this analysis. }  
\label{tab:twolepton}
\end{table}

The bottom panel of Fig.~\ref{Figure:reach_ttHH_2lep} shows the reach of 2 OS $\ell$ signal region combining both the $t\bar{t}hZ\met$ and $t\bar{t}ZZ\met$ channels.   The stop mass up to 930 GeV can be discovered at 5$\sigma$ significance, or a stop mass less than about 1060 GeV is excluded at the 95\% C.L. for the  2 OS $\ell$ signal region.    The specific set of advanced selection cuts used to do the signal combinations are: $\met>150$ GeV, $H_T>$ 500 GeV, $|m_{\ell\ell}-m_Z|<5$ GeV, $m_{T2}>80$ GeV,  $N_j \geq 5$ and $N_{bj}\geq 2$.

\subsection{Results of $\ge 3 \ell$ signal region}

 For  signal region with at least 3 leptons, it is further divided into  ``off-$Z$" and ``on-$Z$" signal region. The ``off-$Z$" signal region is applied to the $t\bar t hh\met$ channel,  while the ``on-$Z$" signal region is applied to the $t\bar t hZ\met$ and $t\bar t ZZ\met$ channels. The cumulative cut efficiencies  after each level of advanced   cuts and cross sections  for the ``on-$Z$" signal region are shown in Table~\ref{tab:threelepton} for the benchmark point.   We do not list such table for the ``off-$Z$" signal region because the reach is very small for all   three channels. As can be seen from Table~\ref{tab:threelepton}, the $t \bar t Z$ is the dominant background, followed by the $t \bar t h$ process. The $t \bar t$ and $t \bar t b \bar b$ processes are highly suppressed.
\begin{table}[tb]
\resizebox{\textwidth}{!}{
 \begin{tabular}{|c|c|c|c|c|c|c|c|c|} 
 \multicolumn{9}{c}{\bf on-$Z$}  \\ \hline
\toprule[1pt]
Process & $\sigma$ (fb)& Basic &$\met>$ & $H_T>$ & $|m_{\ell\ell}-m_Z| <$ &  $N_j\geq$ & $N_b\geq$ & $\sigma$ (fb)\\
        &  & cuts & $175\text{ GeV}$ & $400\text{ GeV}$ & 5 & 4 & 1 & after cuts \\
\toprule[1pt]
\hline
$\tilde{t}_1 \tilde{t}_1 (t\bar t hZ)$ & 80 & 1\% &  0.4\% & 0.3\% & 0.27\% & 0.23\% & 0.23\% & 0.19 \\
\hline
$\tilde{t}_1 \tilde{t}_1 (t\bar t ZZ)$ & 46 & 1.5\% & 0.7\% & 0.5\% & 0.46\% & 0.35\% & 0.35\% & 0.16 \\
\hline
 $t\bar{t}Z$ & 1095 & 0.8\% & $5\times10^{-4}$ & $2\times10^{-4}$ &  $1.7\times10^{-4}$ & $8.1\times10^{-5}$ & $8.1\times10^{-5}$ & 0.09\\
\hline
$t\bar{t}W^{\pm}$ & 747 & $7\times10^{-4}$ & $6.6\times10^{-5}$ & $1.8\times10^{-5}$  & $6.4\times10^{-6} $ & $2.4\times10^{-6}$ & $2.4\times10^{-6}$ & 0.002 \\
\hline
$t\bar{t}h$ & 572 & 0.1\% & $1.1\times10^{-4}$ & $4.4\times10^{-5}$ & $1.6\times10^{-5}$ & $9.4\times10^{-6}$ & $9.4\times10^{-6}$ & 0.005 \\
\hline
$t\bar{t}hh$ & 0.83 & 0.7\% & $9\times10^{-4}$ & $5\times10^{-4}$ & $2\times10^{-4}$ & $1.4\times10^{-4}$ & $1.4 \times 10^{-4}$ & -\\
\hline
$t\bar{t}hZ$ & 1.41 & 1.8\% & 0.21\% & 0.1\% & $9\times10^{-4}$ & $5.7\times10^{-4}$ & $5.7\times10^{-4}$ & -\\
\hline
$t\bar{t}ZZ$ & 1.73 & 2.2\% & 0.32\% & 0.17\% & 0.14\% & $8.3\times10^{-4}$ & $8.3 \times 10^{-4}$ & -\\
 \bottomrule[1pt]
\hline
\end{tabular}
}
\caption{Cumulative cut efficiencies after each level of advanced seletion cuts and cross sections for the signal $\tilde{t}_1\tilde{t}_1^* \rightarrow t \bar thZ\met$ and $t \bar tZZ\met$ as well as SM backgrounds in the $\ge 3 \ell$  ``on-$Z$" signal region at the 14 TeV LHC.  The $t \bar thh$ has no reach sensitivity because of the extremely low branching fraction of three leptons channel. The $t \bar t$ and $t \bar t b \bar b$ processes are also not listed since they are highly suppressed after all the cuts.  }
\label{tab:threelepton}
\end{table}

\begin{figure}[h]
\begin{center}
\includegraphics[width = 0.48 \textwidth]{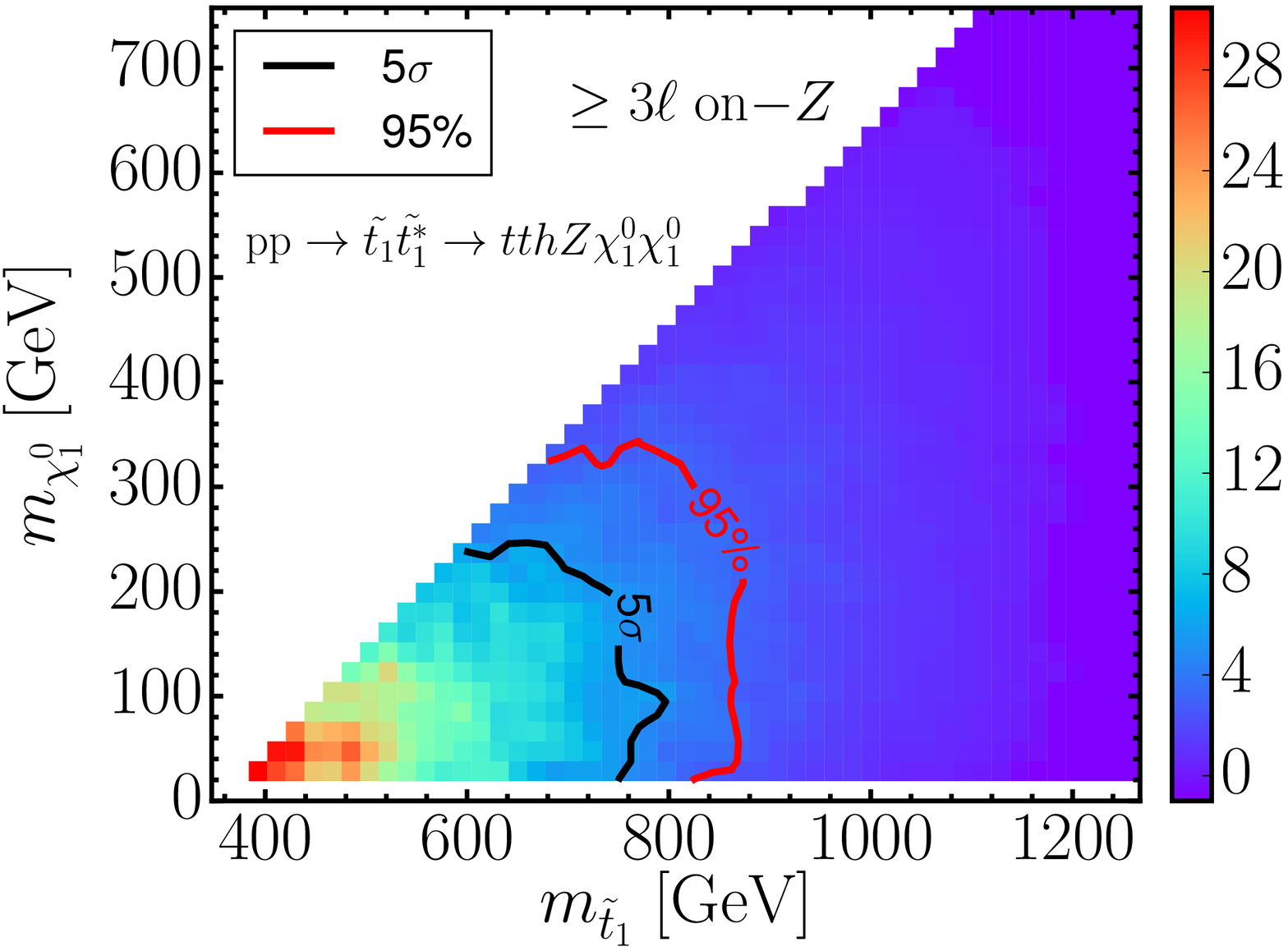}
\includegraphics[width = 0.48 \textwidth]{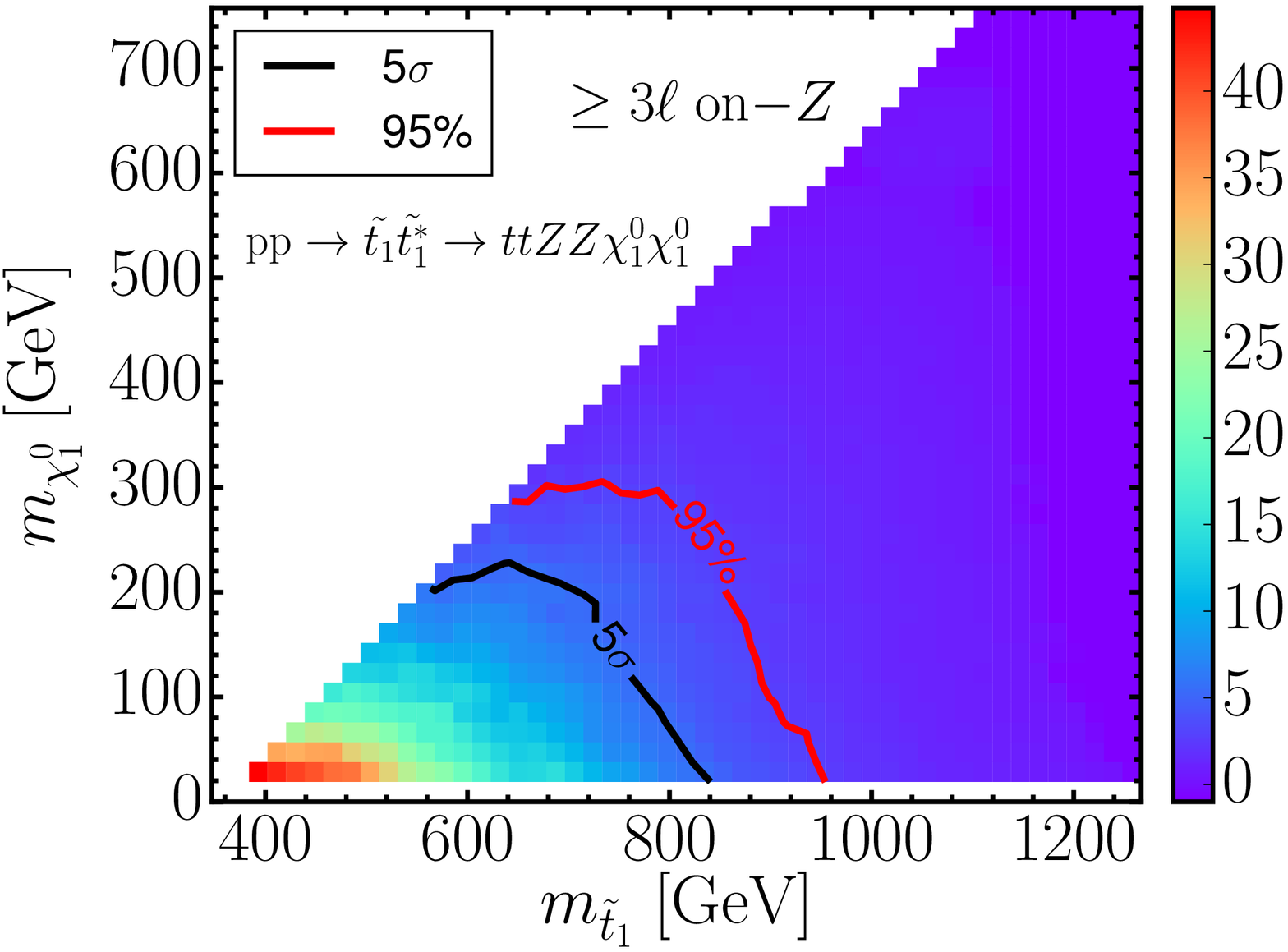}
\includegraphics[width = 0.48 \textwidth]{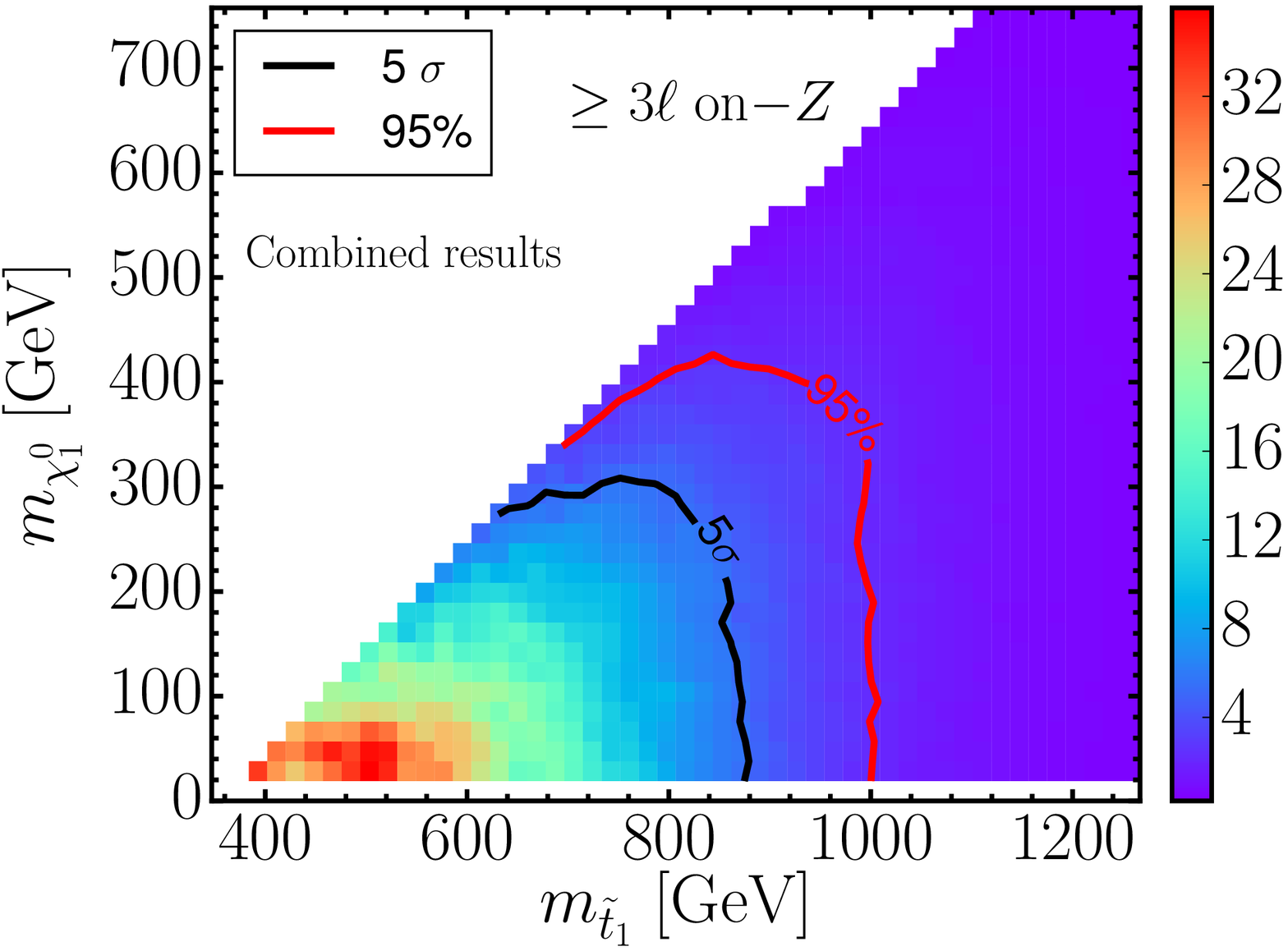}
 \end{center}
\caption[Signal significance contour for at least three leptons channel]{The 95\% CL upper limits (black curve) and 5$\sigma$ discovery reach (red curve) are shown in the plane of MSSM parameter space $m_{\tilde t_1}$ vs $m_{\chi_1^0}$ for the stop pair production $pp \to \tilde t_1 \tilde t_1^* \to t \bar t \chi_2^0/\chi_3^0 \to t \bar t h Z \met$ (top left) and $t \bar t ZZ\met$ (top right) and combined reach (bottom middle)  in the  $\ge 3 \ell$  ``on-$Z$"  signal region at the 14 TeV LHC with 300 ${\rm fb}^{-1}$ integrated luminosity. $\mu$ is fixed to be $M_1$ + 150 GeV.   }
\label{Figure:reach_ttHH_3lep_onZ}
\end{figure}

The 95\% C.L. upper limits (black curve) and 5$\sigma$ discovery reach (red curve) for the ``on-$Z$" signal region are shown in Fig.~\ref{Figure:reach_ttHH_3lep_onZ}.   10\%  systematic uncertainties are assumed.  $t\bar{t}hh\met$ channel has almost no reach, therefore not shown in the plot.    A stop mass up to 780 GeV (850 GeV) for the channel $t \bar t hZ\met$ ($t \bar t ZZ\met$) can be discovered at the 5$\sigma$ significance, and up to about  860 GeV (960 GeV)  for 95\% C.L. exclusion.  Limits with 20\% systematic uncertainties are very similar to that of 10\% case.  The reach for the ``off-$Z$" signal region is much smaller than that of ``on-$Z$" signal region.

The combined reaches of $t\bar{t}hZ\met$ and $t\bar{t}ZZ\met$ channels for the $\ge 3 \ell$ 
 ``on-$Z$" signal region are shown in the bottom panel of Fig.~\ref{Figure:reach_ttHH_3lep_onZ}.
The 5$\sigma$ reach of a stop mass is about   880 GeV, and the 95\% C.L. exclusion limit can reach up to 1000 GeV.
The specific set of advanced selection cuts used to do the signal combinations are: $\met>200$ GeV, $H_T>$ 500 GeV,  $|m_{\ell\ell}-m_Z|<5$ GeV ,  $N_j \geq 7$ and $N_{bj}\geq 1$.

\subsection{Results of combined channels}

\begin{figure}[!htbp]
\begin{center}
\includegraphics[width = 0.48 \textwidth]{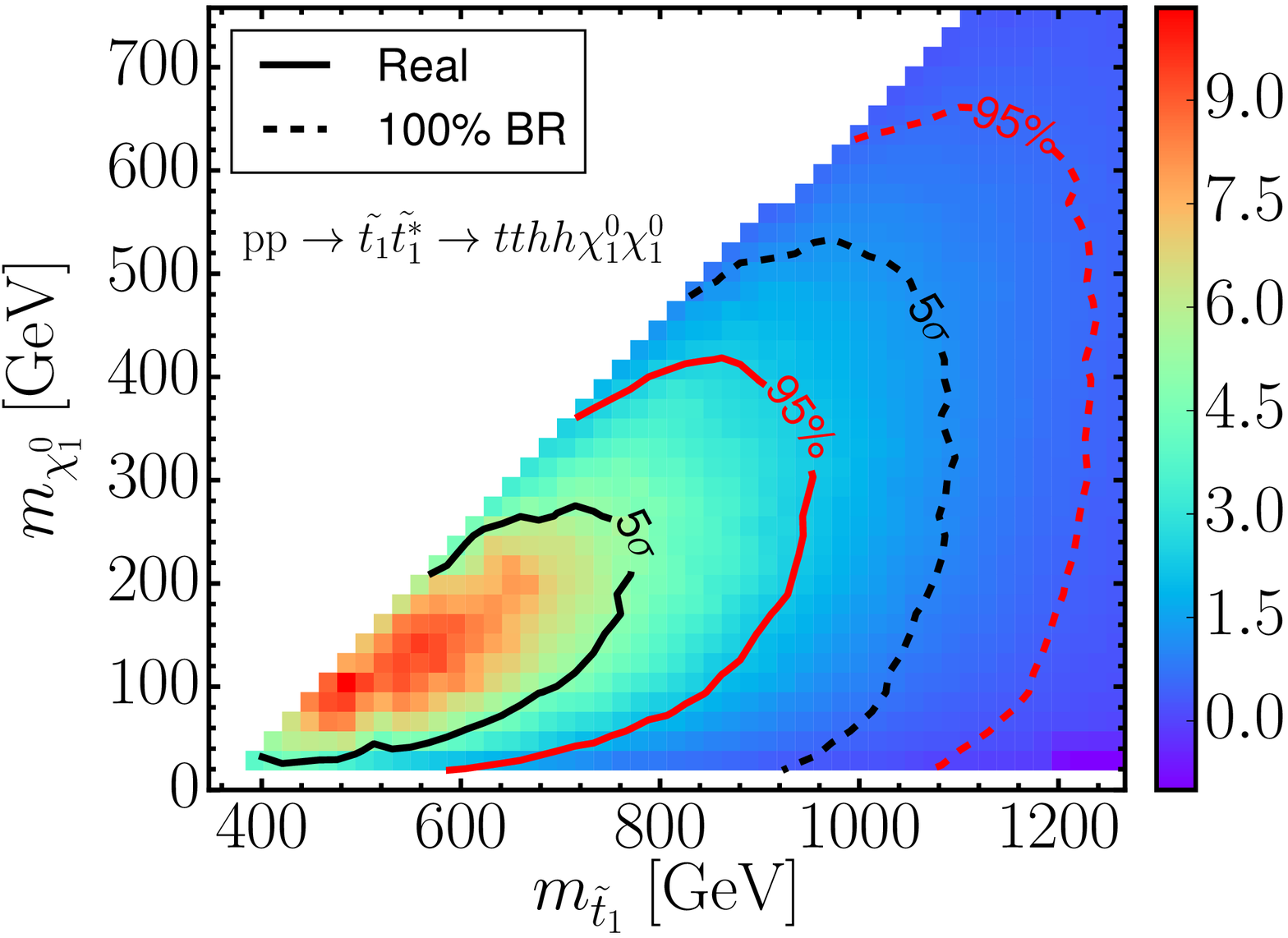}
\includegraphics[width = 0.48 \textwidth]{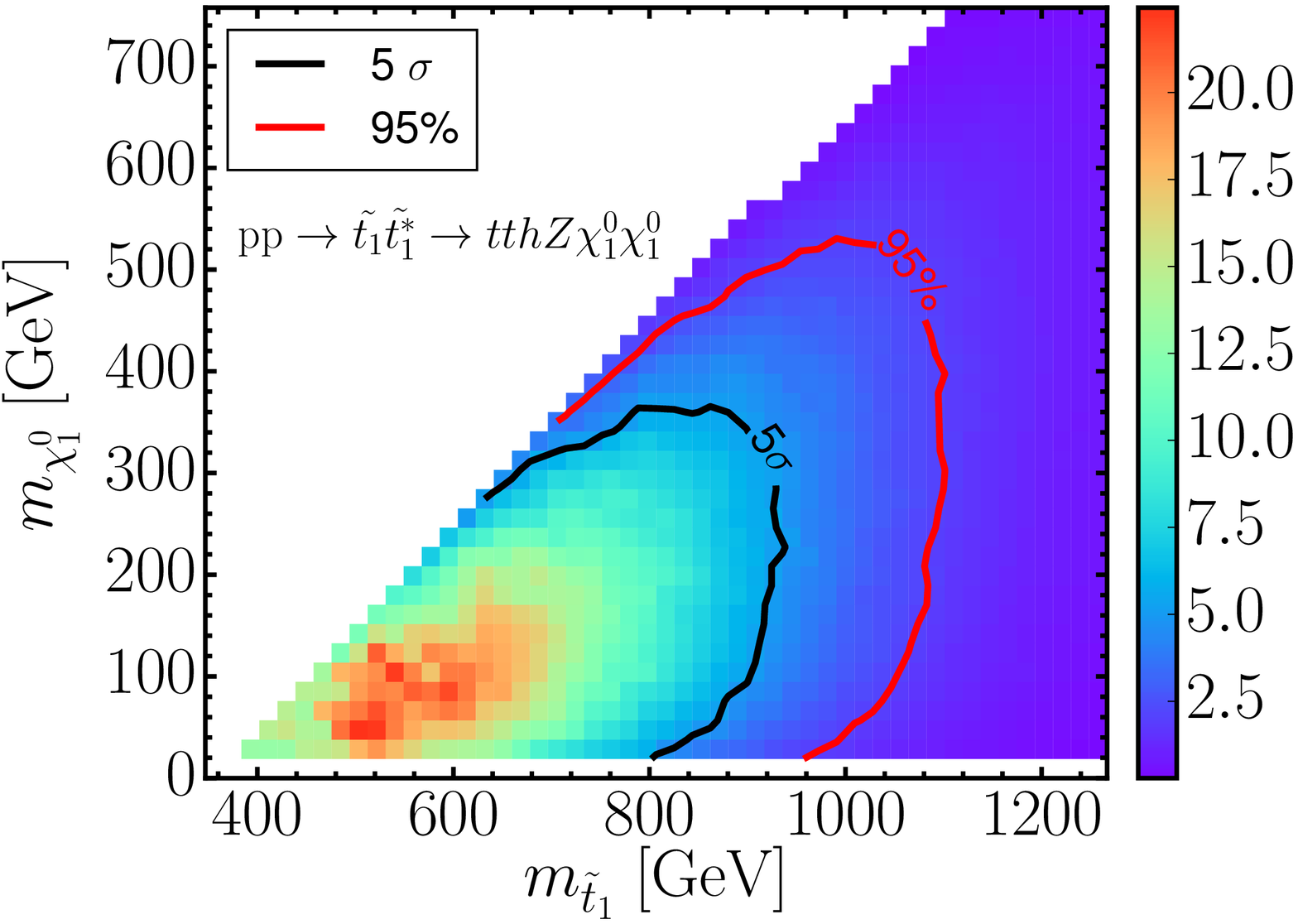}
\includegraphics[width = 0.48 \textwidth]{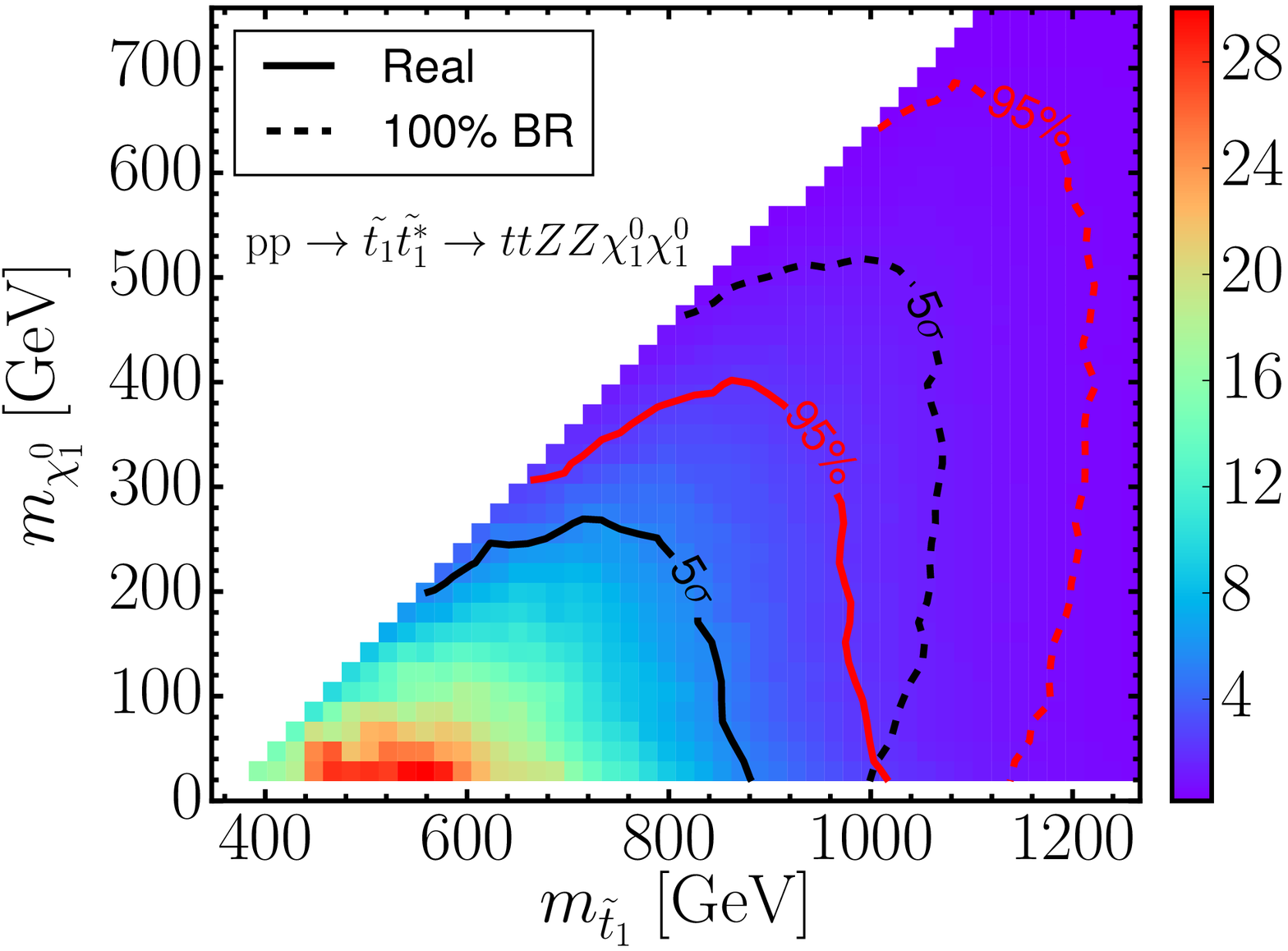}
 \end{center}
	\caption[Signal significance contour for the combined signal regions]{The 95\% CL upper limits (black curve) and 5$\sigma$ discovery reach (red curve)  for the combination of three  signal regions for  the  $t\bar{t}hh\met$ (top left panel), $t \bar t hZ\met$  channel (top right panel), $t \bar t ZZ\met$ channel (bottom panel) at the 14 TeV LHC with 300 ${\rm fb}^{-1}$ integrated luminosity.     The dotted line in the top left (bottom)  panel represents the reach of the channel $t \bar t hh\met$ ($t \bar t ZZ\met$) assuming a 100\% decay branching fraction.  }
\label{Figure:reach_com1}
\end{figure}

For each signal region, the combined reach of all three channels are shown in previous sections. Here we discuss the reach of each individual channel, combining all the signal regions.  In Fig.~\ref{Figure:reach_com1}, we show the 5$\sigma$ discovery reach (red curve) and 95\% C.L. exclusion limit (black curve) for combination of three signal regions of $t\bar{t}hh\met$ (top left panel), $t\bar{t}hZ\met$ channel (top right panel) and $t\bar{t}ZZ\met$ channel (bottom middle panel).  Since both the 2 OS $\ell$ and $\ge 3 \ell$ have no reach for $t\bar{t}hh\met$ channel, the combined reach for $t\bar{t}hh\met$ is simply the 1 $\ell$ reach as in the top left panel of Fig.~\ref{Figure:reach_1lep}.   For both $t\bar{t}hZ\met$ and $t\bar{t}ZZ\met$ channel, 1 $\ell$ signal region gives the best sensitivity.   The 5$\sigma$ reach can discover a stop with mass up to 950 GeV,  and the 95\% exclusion limits can reach up to 1100 GeV for the channel $t \bar t hZ\met$. The corresponding limit for the channel $t \bar t ZZ\met$ is a little weaker due to the smaller branching fraction.  In dashed lines, we also show the reach of $t\bar{t}hh\met$ and $t\bar{t}ZZ\met$ channel assuming a 100\% decay branching fraction.

\section{Collider analysis at $\sqrt{s} =$ 100 TeV}
\label{sec:analysesof100}

To explore the physics potential of  the future 100 TeV $pp$ machine, it is critical to explore the complete parameter space of the MSSM.  We scan the MSSM stop and neutralino/chargino mass parameter in the following region:   

$\bullet$
$M_{3SQ}$ from 1000 to 8000 GeV in a step of 250 GeV, corresponding to $m_{\tilde{t}_1}$ from 1009 GeV to  8001 GeV.

$\bullet$
$M_1$ is scanned from 5 GeV to 5000 GeV,  in the step of 250 GeV.

$\bullet$
$\mu$ is fixed to be $\mu$ = $M_1$ + 500 GeV.

$\bullet$
We further require $m_{\tilde{t}_1}>m_{\chi_2^0}/m_{\chi_3^0}+m_t$ such that $\tilde{t}_1\to t\chi_2^0/\chi_3^0$  is kinematically open.

At the 100 TeV future machine, the decay kinematics will be significantly different from that of the LHC. The decay products such as the top quark from heavy stop are highly boosted as discussed in Ref. \cite{Cohen:2014hxa}, leading to highly collinear leptons with the high $p_T$ jets. So we do not require the separation $\Delta R(j,l)$ between jets and leptons to be larger than 0.5 at the Monte Carlo event generation stage. The Delphes 3 Snowmass combined LHC No-Pile-up detector card \cite{Anderson:2013kxz} is modified for the 100 TeV future collider for the detector simulation.   We allow up to one additional parton in the final state, and adopt the MLM matching scheme \cite{Mangano:2006rw} with  xqcut = 80 GeV for $t \bar t j$ background. Both the SM backgrounds and the stop pair production signal are normalized to theoretical cross sections, calculated including higher-order QCD corrections \cite{Borschensky:2014cia, Avetisyan:2013onh}.  At the event generation level, we apply the $S_T$ cut (the scalar sum of $p_T$ for all partons) as following: $S_T \ge$ 3 TeV for the  $t \bar t j$ background and $S_T \ge$ 1 TeV for the $t \bar t B$ background, where $B$ stands for bosons including $W$, $Z$ and $h$.

We apply the following   cuts for both the signal and the SM backgrounds:

\begin{itemize}
\item{All jets reconstructed using anti-$k_t$ algorithm \cite{Cacciari:2008gp} with cone radius $R=0.5$  are required to have $p_T >$ 50 GeV and $|\eta| <$ 2.5, including  at least two jets with $p_T >$ 1000 (500) GeV.  }
\item{All leptons ($e$ or $\mu$) are required to have $p_T >$ 30 GeV and $|\eta| <$ 2.5, including at least one lepton with $p_T >$ 100 (200) GeV contained within a $ \Delta R$ = 0.5 cone centered around one of the two leading   jets.  }
 \item{The  separation $\Delta \Phi({\bf p}_T^{miss},j)$ between the missing transverse momentum and jets with $p_T >$ 100 (200) GeV and $|\eta| <$ 2.5 is required to be larger than 1.0. }
\item{$m_T$  to be greater than 500, 750, 1000, 1250, 1500 GeV. }
\item{$\met$ to be greater than 1, 1.5, 2, 2.5, 3, 3.5, 4 TeV.}
\item{$H_T$ to be greater than 2, 3, 4, 5, 6 TeV.}
\item{$N_j$ to be at least 4, 5, 6, 7; $N_{bj}$ to be at least 2, 3 ,4 ,5.}

\end{itemize}
\begin{figure}[!htbp]
\begin{center}
\includegraphics[width = 0.48 \textwidth]{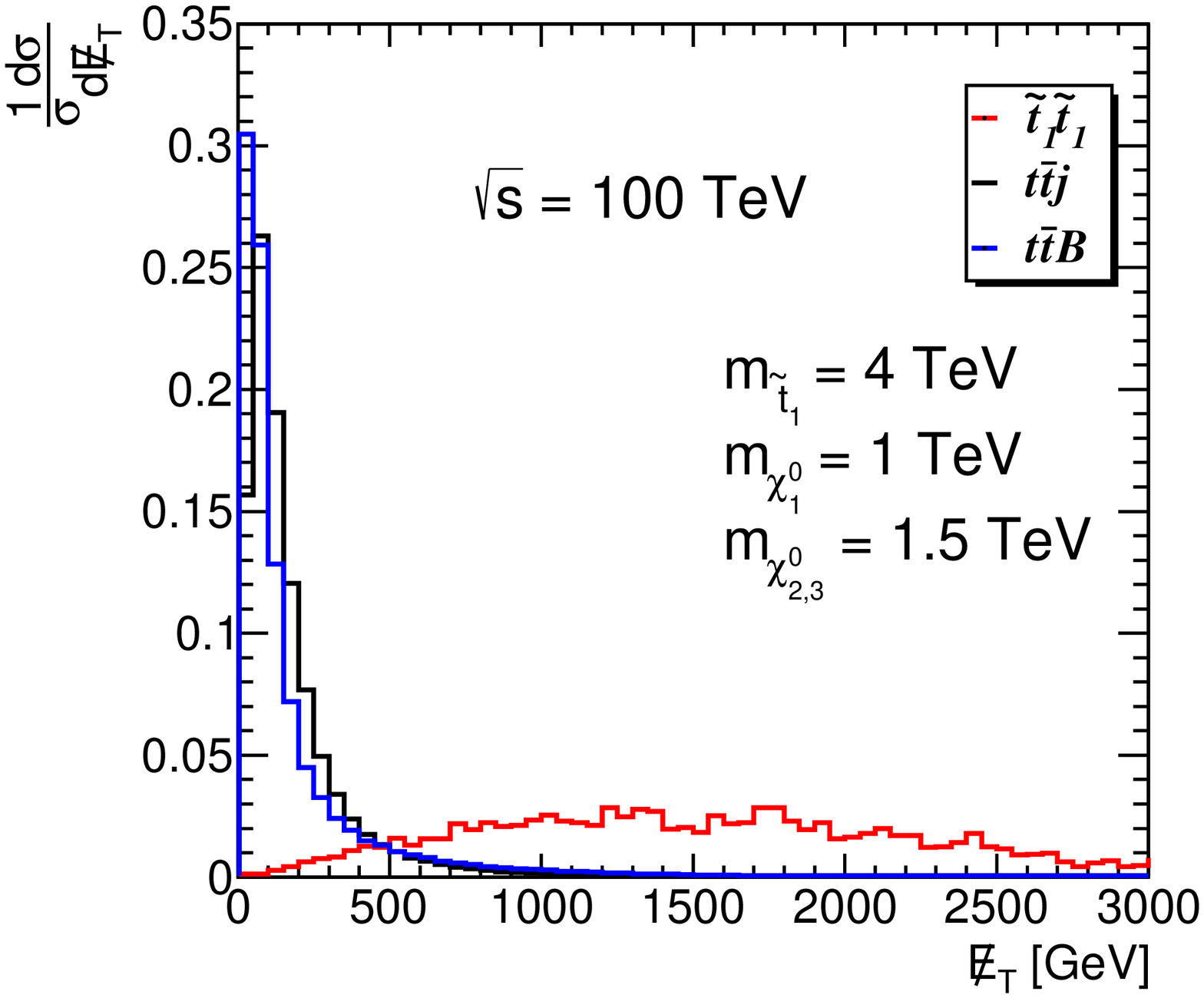}
\includegraphics[width = 0.48 \textwidth]{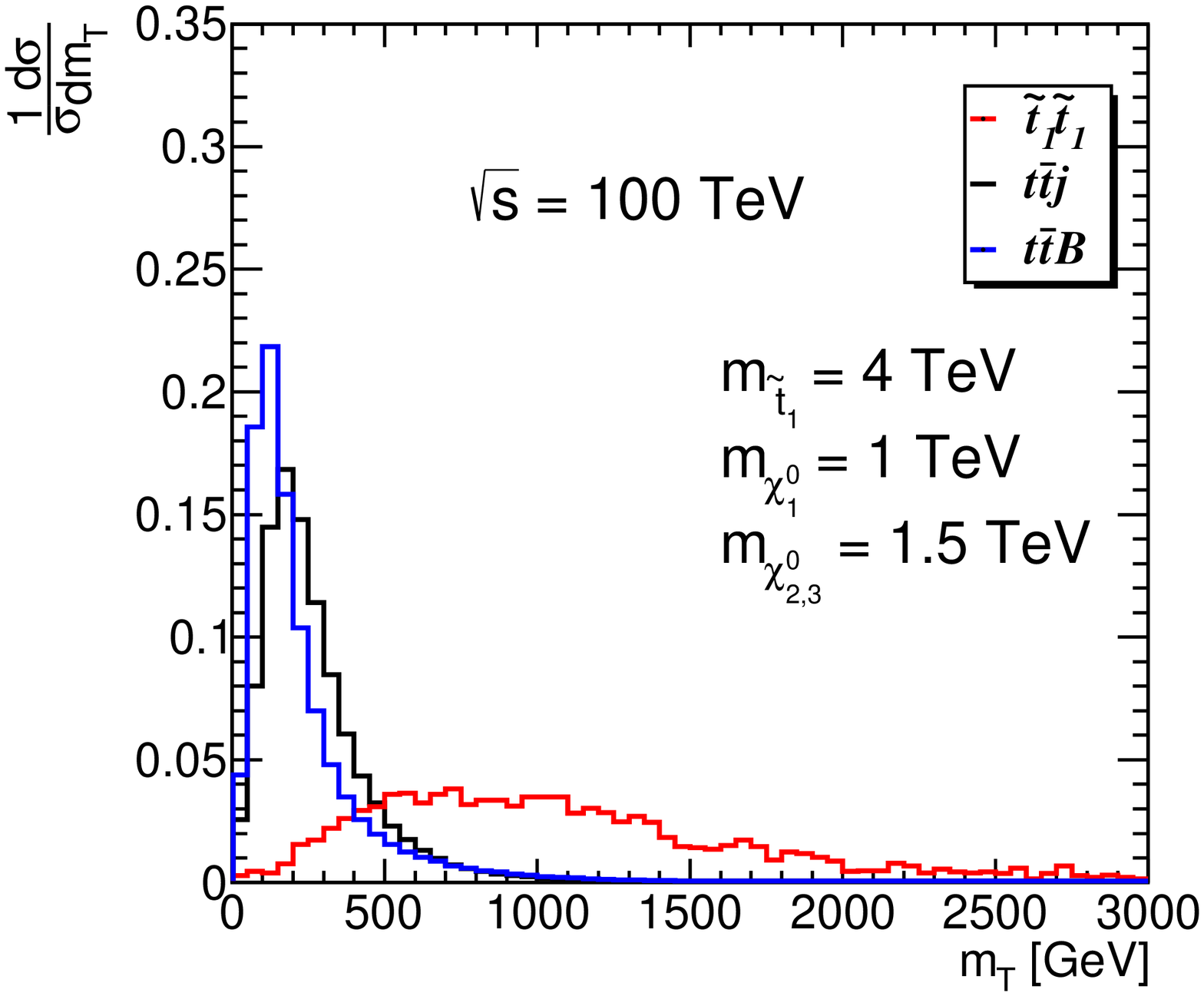}
\end{center}
	\caption[Normalized distributions of $\met$ and $m_T$ at $\sqrt{s}$ = 100 TeV]{Normalized distributions of $\met$ (left) and $m_T$ (right) for the signal channel $t \bar t hZ\met$  with $m_{\tilde t_1} =$ 4000 GeV and $m_{\chi_1^0} =$ 1000 GeV as well as SM backgrounds after the $N_j$, $N_\ell$ and  $\Delta \Phi({\bf p}_T^{miss},j)$ cuts. }
\label{Figure:plots_100TeV}
\end{figure}

The above selection cuts are efficient to suppress the SM backgrounds. For example, after imposing the collinear  leptons   to the two leading jets requirement on the SM backgrounds, the selected samples mainly contain the boosted heavy quarks. The neutrinos in the form of $\met$ from their decay are highly aligned with the jet momenta. However, the signal $\met$ has extra contribution from the LSP, which is usually not aligned with the jet momenta. Therefore it is useful to impose the angle separation $\Delta \Phi({\bf p}_T^{miss},j)$ cut between $\met$ and the jets with $p_T^j >$ 100 (200) GeV and $|\eta|^j <$ 2.5 to suppress the $t \bar tj$ and $t \bar tB$ backgrounds. The normalized distributions of $\met$ and $m_T$ after the above cuts are displayed in Fig.~\ref{Figure:plots_100TeV}. The $\met$ and $m_T$ distributions of the signal are very broad because of the extra contribution from the LSP. Contrarily, the $\met$ and $m_T$ distributions of the SM backgrounds are typically bounded around $m_W$. Those two selection cuts are highly efficient to suppress  the SM backgrounds.   Table~\ref{tab:100TeV} shows the  cross sections, yields and cumulative cut efficiencies after each level of selection cut for the signals with $m_{\tilde t_1} =$ 4000 GeV and $m_{\chi_1^0} =$ 1000 GeV  as well as the SM backgrounds.  The $t \bar t B$ ($B = W, Z, h$) is the dominant background after all cuts. Other SM backgrounds are typically small after strong selection cuts, then they can be neglected. The discovery significance for the channel $t \bar t hZ\met$ can reach 7$\sigma$ for this benchmark point.

\begin{table}[h]
\centering
 
     \begin{tabular}{|c|c|c|c|c|c|}
\toprule[1pt]
    Description & $\tilde{t}_1 \tilde{t}_1$ ($t \bar t hh$) & $\tilde{t}_1 \tilde{t}_1$ ($t \bar t hZ$)  & $\tilde{t}_1 \tilde{t}_1$ ($t \bar t ZZ$)  & $t\bar t j$ & $t\bar t B$  \\
\toprule[1pt]
    CS (fb) before cuts & 0.66 & 1.32 & 0.67 & 2670 & 2003  \\ \hline
    $N_j \ge$ 2 & 94\% & 93\% & 92\% & 93\% & 40\% \\ \hline
    $N_\ell \ge$ 1 & 37\% & 37\% & 35\% & 40.6\% & 8.6\% \\ \hline
    $\Delta \Phi(j,\met)$ & 5.5\% & 6.6\% & 7.5\% & 1.7\% & $7.3\times 10^{-3}$ \\ \hline
    $\met >$ 1500 GeV & 2.5\% & 3.3\% & 4.1\% & $2.9\times 10^{-5}$ & $6.6\times 10^{-5}$  \\ \hline
    $H_T >$ 4000 GeV & 1.2\% & 1.6\% & 1.9\% & $1.0\times10^{-5}$ & $8.5\times 10^{-6}$ \\ \hline
    $m_T >$ 1000 GeV & $7.4\times 10^{-3}$ & 1.1\% & 1.4\% & $3.2\times10^{-6}$ & $5.4\times10^{-6}$ \\ \hline
    $N_j \ge$ 5 & $5.8\times 10^{-3}$ & $8.3\times 10^{-3}$ & $8.7\times 10^{-3}$ & $1.8\times10^{-6}$ & $2.4\times10^{-6}$ \\ \hline
    $N_{bj} \ge$ 2 & $4.9\times 10^{-3}$ & $6.0\times 10^{-3}$ & $5.5\times 10^{-3}$ & $3.1\times10^{-7}$ & $1.3\times10^{-6}$ \\ \hline
    CS (fb) after cuts & $3.2\times 10^{-3}$ & $8.0\times 10^{-3}$ & $3.7\times 10^{-3}$ & $8.3\times10^{-4}$ &  $2.6\times 10^{-3}$ \\ \hline
    Event Yields  (3${\rm ab}^{-1}$)& 9.6 & 24 & 11.1 & 2.5 & 7.8 \\
\bottomrule[1pt]
  \end{tabular}

\caption[Cut efficiencies, cross sections and yields for the signal as well as SM backgrounds at $\sqrt{s}$ = 100 TeV]{The cumulative cut efficiencies, cross sections and yields for the signal with $m_{\tilde t_1} =$ 4000 GeV and $m_{\chi_1^0} =$ 1000 GeV as well as SM backgrounds for 100 TeV $pp$ collider with 3 ${\rm ab}^{-1}$ integrated luminosity.   The $B$ stands for bosons including $W$, $Z$ and $h$.  }
\label{tab:100TeV}
\end{table}

\begin{figure}[!htbp]
\begin{center}
\includegraphics[width = 0.48 \textwidth]{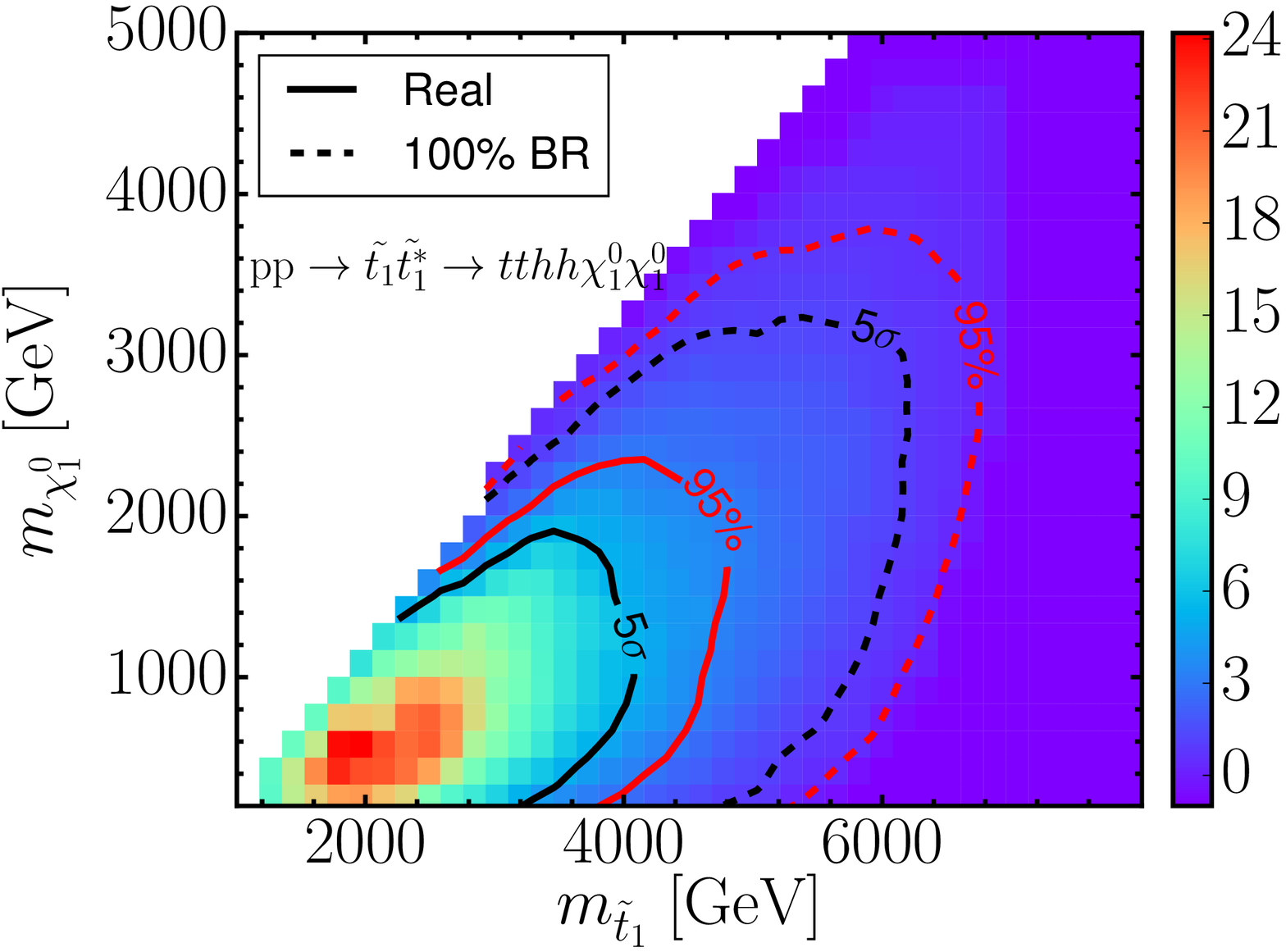}
\includegraphics[width = 0.48 \textwidth]{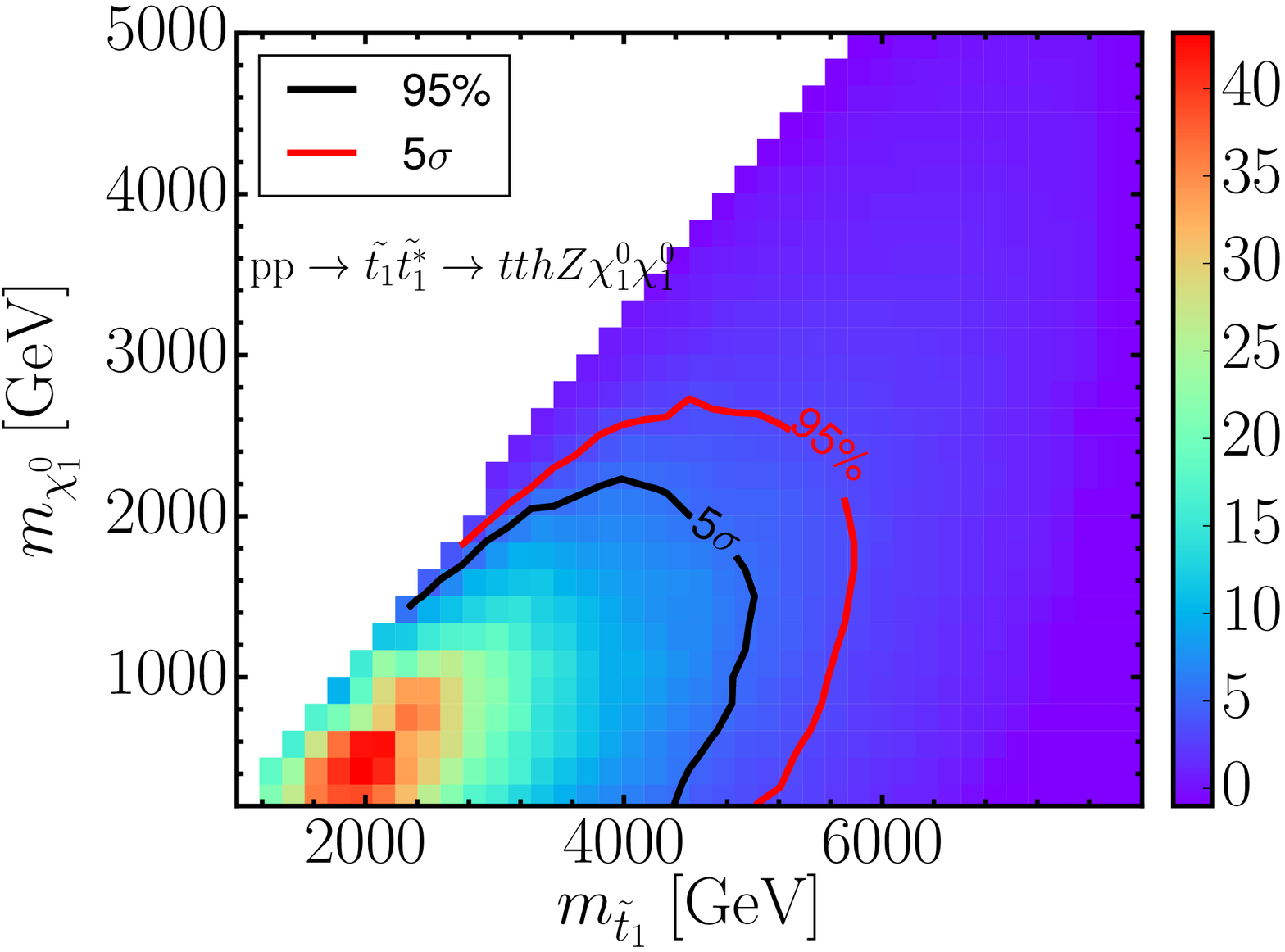}
\includegraphics[width = 0.48 \textwidth]{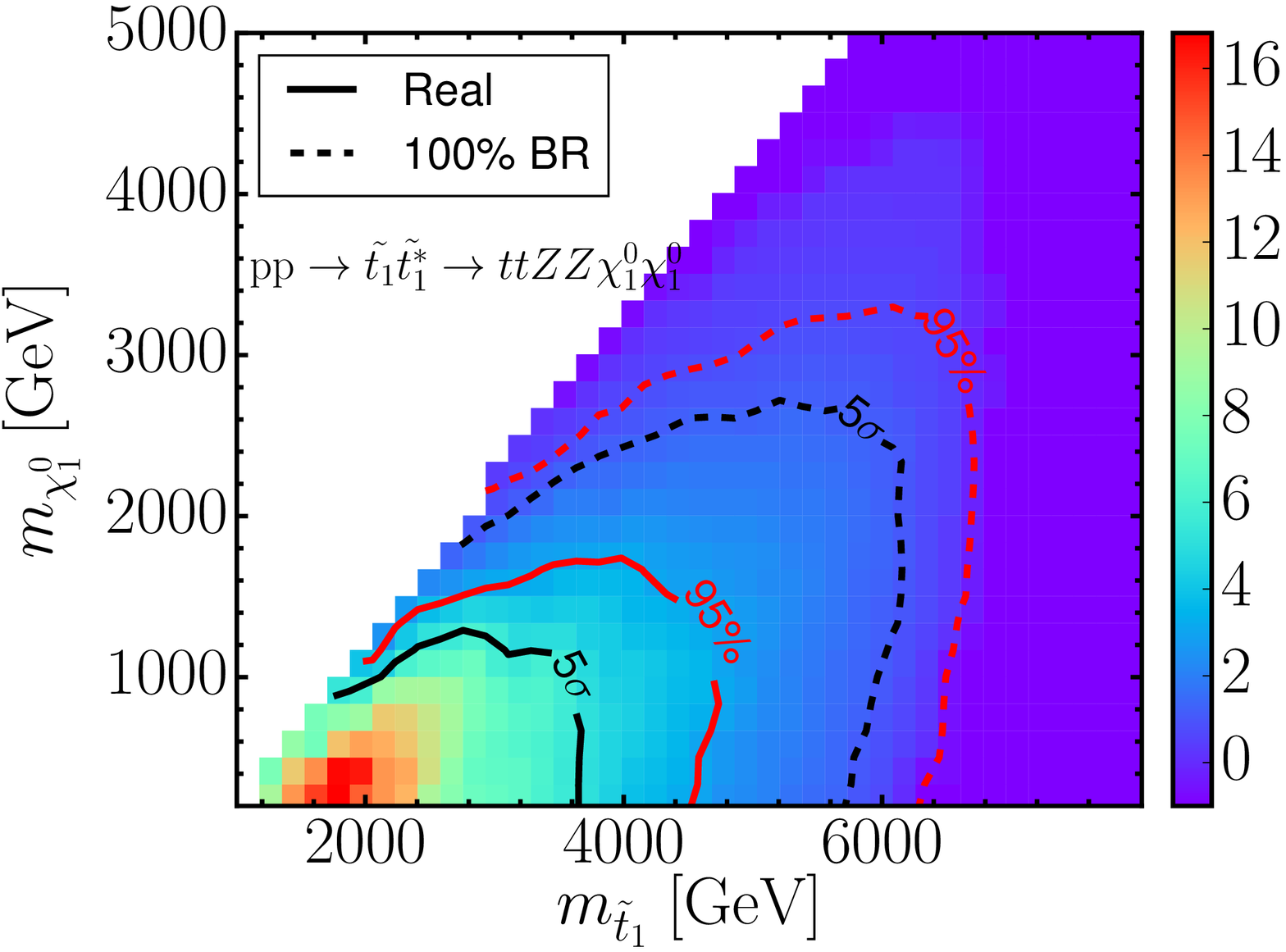}
\includegraphics[width = 0.48 \textwidth]{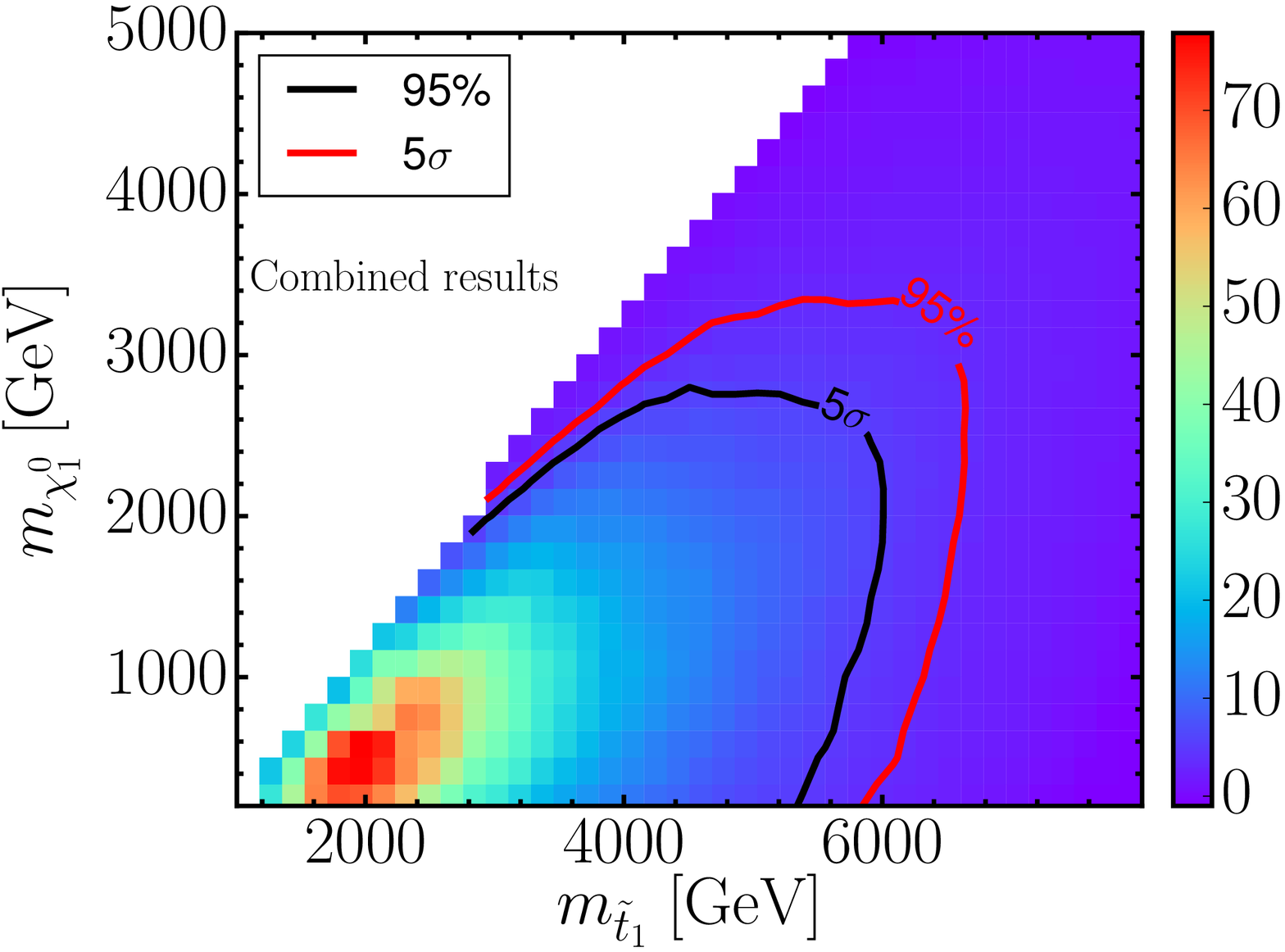}
\end{center}
	\caption[Signal significance contour for three primary channels as well as combined channels at $\sqrt{s}$ = 100 TeV]{The 95\% CL upper limits (black curve) and 5$\sigma$ discovery reach (red curve) are shown in the plane of MSSM parameter space $m_{\tilde t_1}$ vs $m_{\chi_1^0}$ for the stop pair production $pp \to \tilde t_1 \tilde t_1^* \to t \bar t \chi_2^0/\chi_3^0 \to t \bar t h h \met$  (top left), $t \bar t hZ\met$ (top right), $t \bar t ZZ\met$ (bottom left) and combined channels (bottom right)  at 100 TeV future $pp$ collider with 3000 ${\rm fb}^{-1}$ integrated luminosity  for $\geq 1 \ell$ signal region. $\mu$ is fixed to be $M_1$ + 500 GeV and 10\% systematic uncertainties are assumed. Solid line stands for the realistic MSSM scenario and dotted line represents the simplified model assuming a 100\% decay branching fraction. The color coding on the right indicates the signal significance  to guide the eye. }
\label{Figure:reach_100TeV}
\end{figure}

In Fig.~\ref{Figure:reach_100TeV}, the 95\% C.L. upper limits (black curve) and 5$\sigma$ discovery reach (red curve) based on  $\geq 1 \ell$ signal regions    are shown in the plane of MSSM parameter space $m_{\tilde t_1}$ vs $m_{\chi_1^0}$ for the stop pair production $pp \to \tilde t_1 \tilde t_1^* \to t \bar t \chi_2^0/\chi_3^0 \to t \bar t h h \met$ (top left), $t \bar t hZ\met$ (top right), $t \bar t ZZ\met$ (bottom left) and all    channels  combined (bottom right) at 100 TeV LHC with 3 ${\rm ab}^{-1}$ integrated luminosity.   $\mu$ is fixed to be $M_1$ + 500 GeV and 10\% systematic uncertainties are assumed.     The channel $t \bar t hZ\met$ has the best reach sensitivity due to its large branching fraction, with discovery reach about  5 TeV and exclusion reach about 6 TeV.   Combining all three channels, the discovery (exclusion) reach could be pushed    to about 6 (6.6) TeV.  This will greatly improve our understanding of the TeV scale SUSY and the nature of electroweak breaking.   For $t\bar{t}hh\met$ and $t\bar{t}ZZ\met$, we also show the reach assuming a 100\% decay branching fraction in dashed lines.

\section{Summary and Conclusion}
\label{sec:conclusion}

Most of the current stop searches at the LHC have been performed considering the channels of $t\bar{t}\met$, $bbWW\met$ for the stop sector,  assuming the stop 100\% decaying to either $t\chi_1^0$ or $b \chi_1^\pm$. However, in   MSSM parameter space with light neutralinos and charginos other than the LSP, these decay channels become subdominant or even highly suppressed, resulting in much relaxed bounds from current LHC searches.  In this work, we studied the stop decay behavior in the scenario of a Bino-like LSP ($M_1$) with Higgsino-like NLSPs ($\mu$).  The new decay channels of $\tilde{t}_1\rightarrow t \chi_2^0/\chi_3^0$ dominate because of the large ${\rm SU}(2)_L$ coupling and top Yukawa coupling. Given the further decays of $\chi_2^0/\chi_3^0$ to a Higgs boson or $Z$ boson, the stop pair production at the LHC leads to $t \bar thh\met$, $t \bar thZ\met$ and $t \bar tZZ\met$ final states.

\begin{figure}[!htbp]
\begin{center}
\includegraphics[width = 0.5 \textwidth]{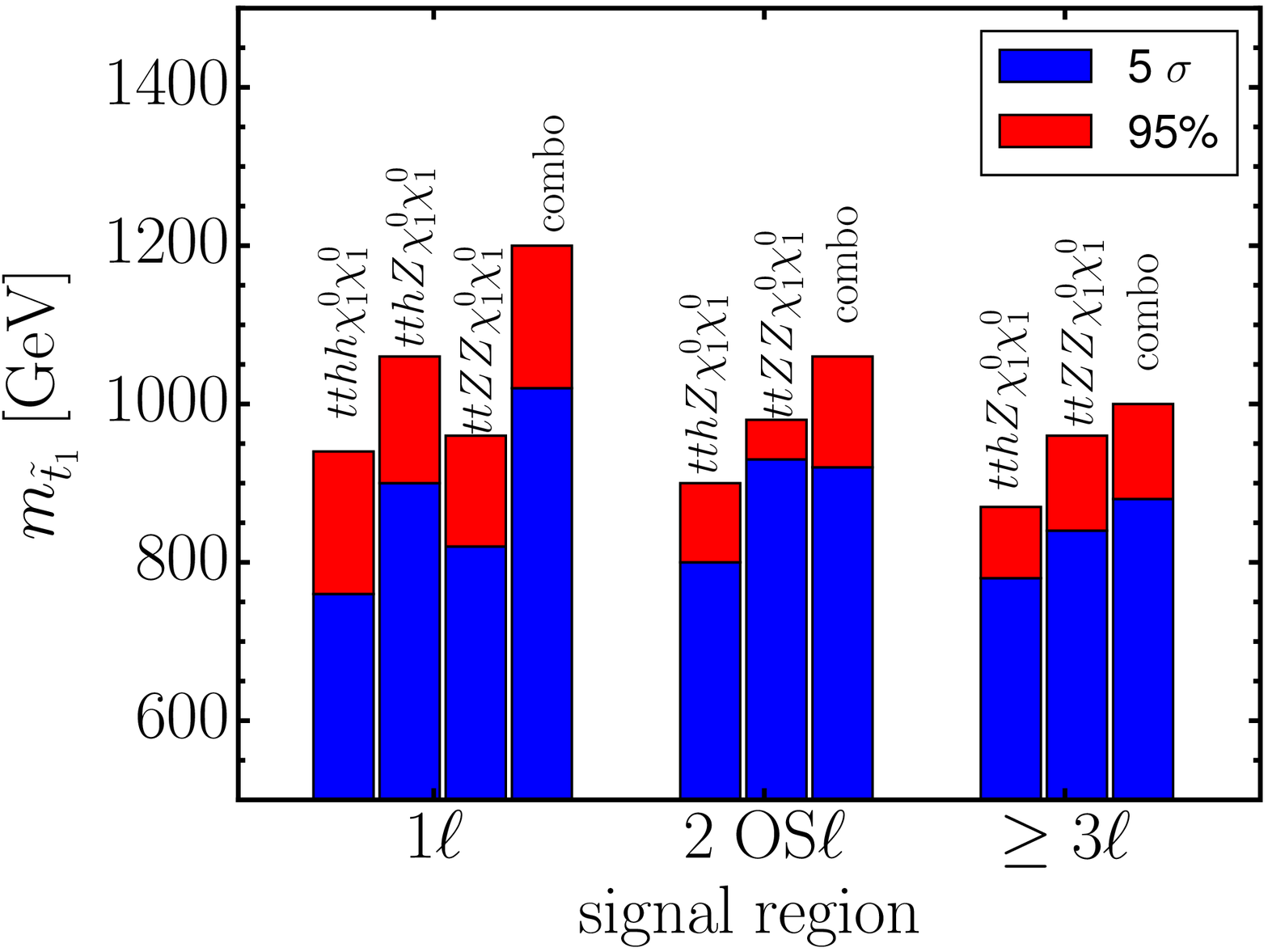}
\includegraphics[width = 0.46 \textwidth]{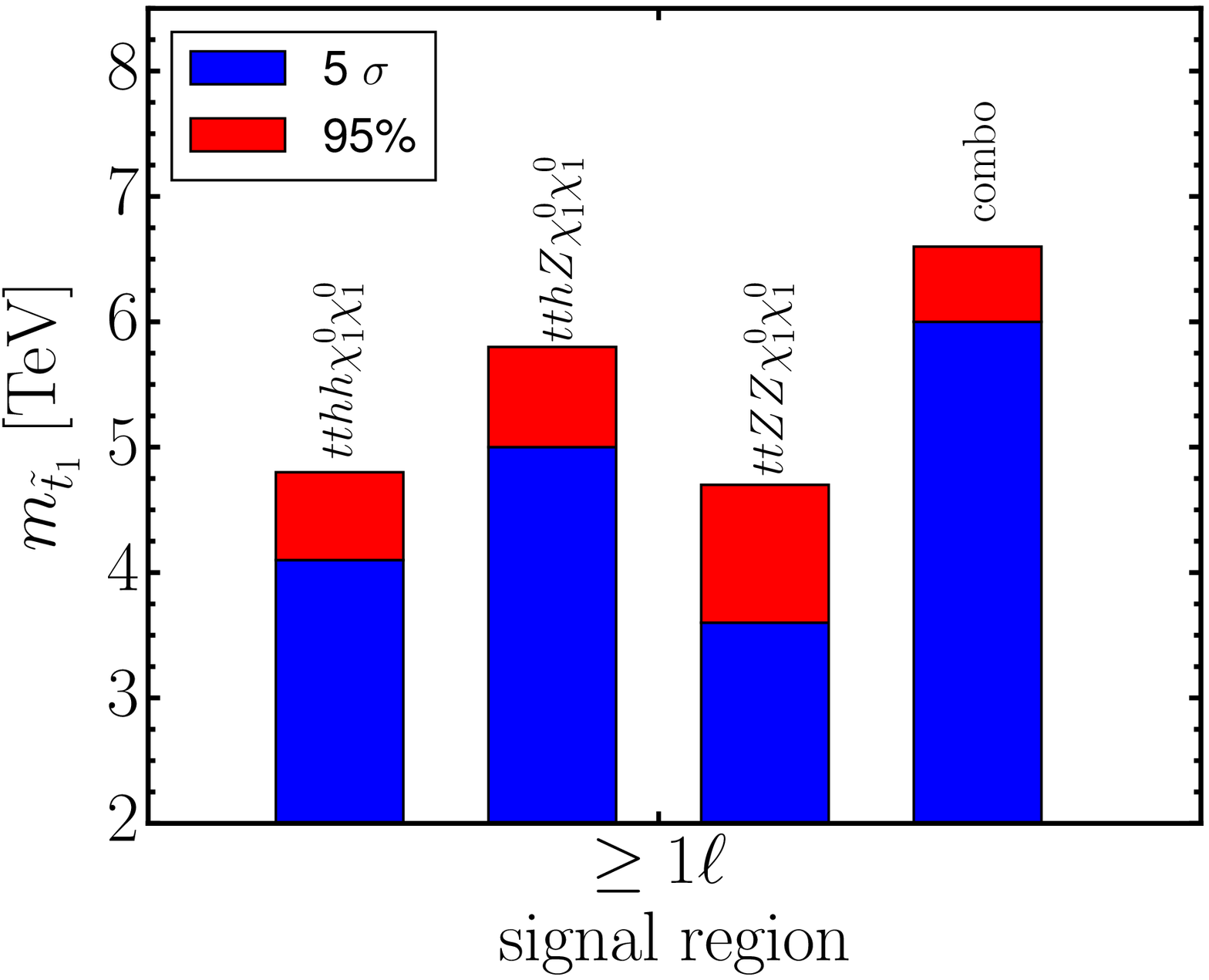}
\end{center}
	\caption[]{The 5 $\sigma$ discovery reach and 95\% exclusion limit of light stop mass for three primary signal regions at the 14 TeV LHC with 300 ${\rm fb}^{-1}$ integrated luminosity (left panel) and at least one lepton signal region at the future 100 TeV $pp$ collider   with 3000 ${\rm fb}^{-1}$ integrated luminosity (right panel).}
	\label{Figure:summary}
\end{figure}

In this work, we focused on the stop search sensitivity at the 14 TeV LHC with 300 ${\rm fb}^{-1}$ integrated luminosity,   in three primary signal regions based on lepton multiplicities: 1 $\ell$, 2 OS $\ell$ and $\ge 3 \ell$.    We combined all the three production channels or three signal regions to obtain the best reach.        We also explore the reach at the future 100 TeV $pp$ collider with 3000 ${\rm fb}^{-1}$ integrated luminosity.    The 95\% C.L. exclusion and 5 $\sigma$ discovery reach are summarized in Fig.~\ref{Figure:summary}.

Although we only consider one very interesting scenario of MSSM parameter space, it is important to identify the leading decay channels in various regions of parameter space to fully explore the reach of the LHC for the third generation squarks, which has important implications for the stabilization of the electroweak scale in supersymmetric models. The strategy developed in our analyses can be applied to the study of top partners in other new physics scenarios as well.

\acknowledgments
We thank Yongcheng Wu and Fionnbarr O'Grady for helpful discussion. The work is supported by the  Department of Energy  under Grant~DE-FG02-04ER-41298.   

\bibliography{bibliography}

\end{document}